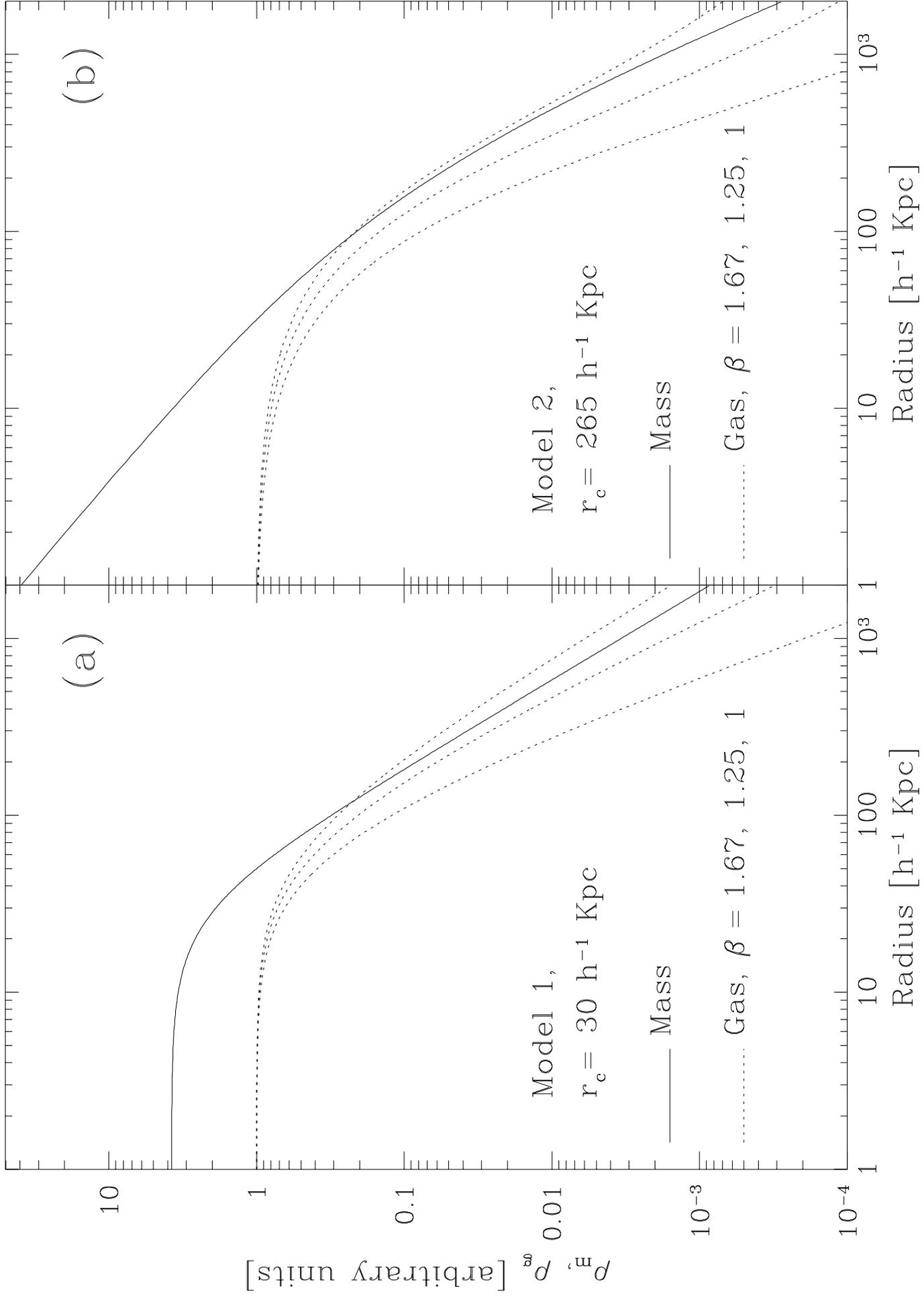

Fig 1

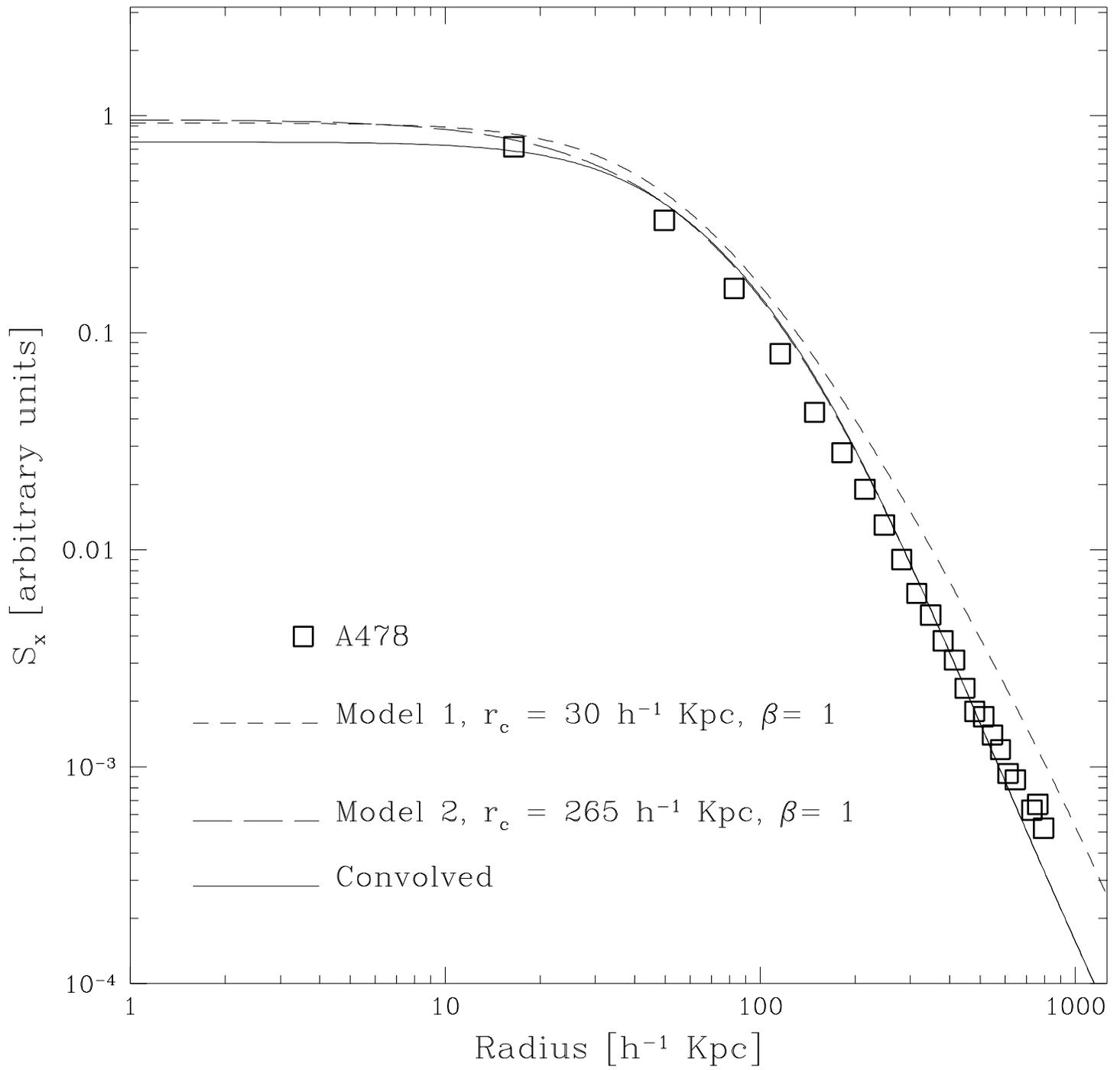

Fig. 2

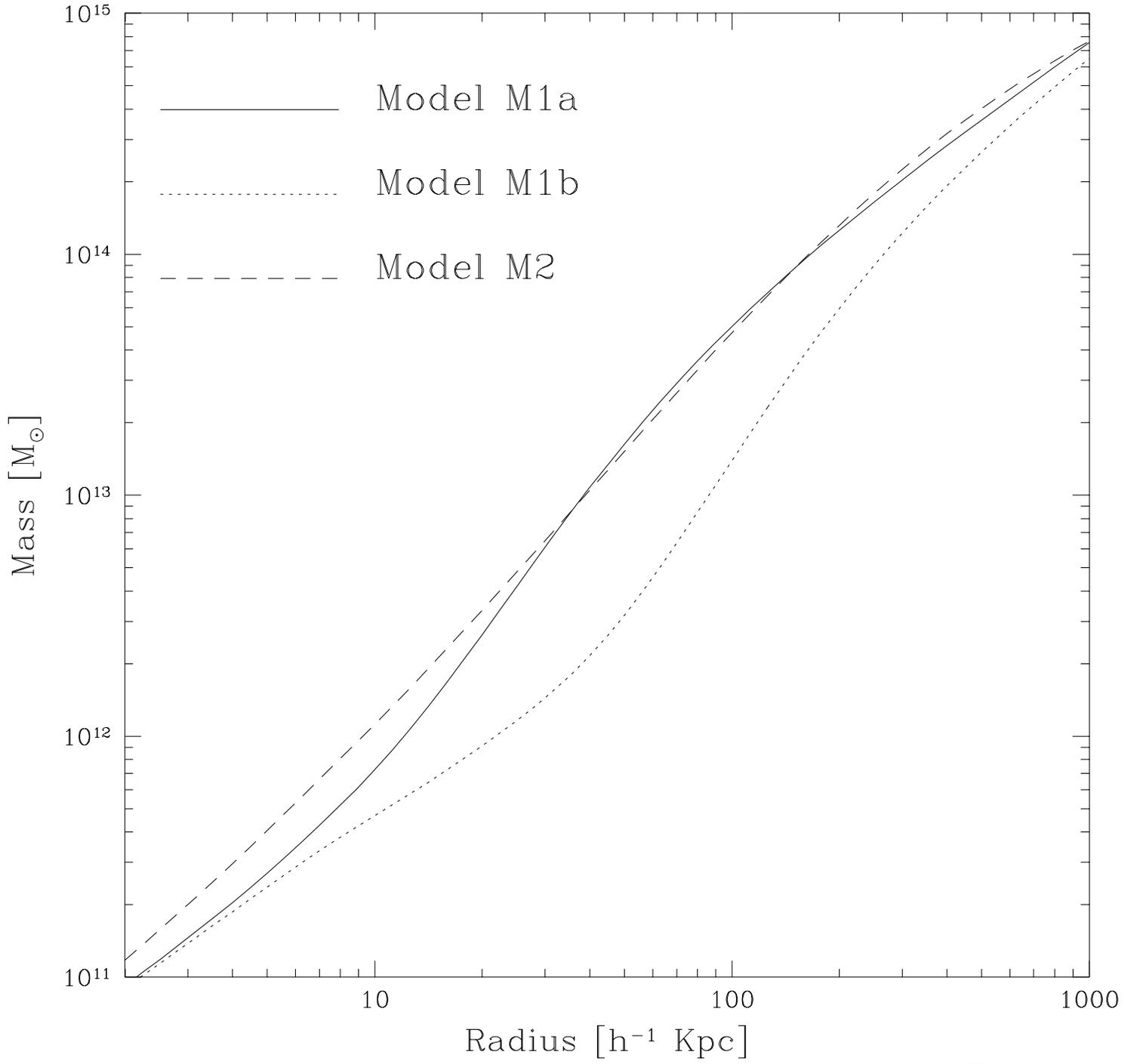

Fig. 3

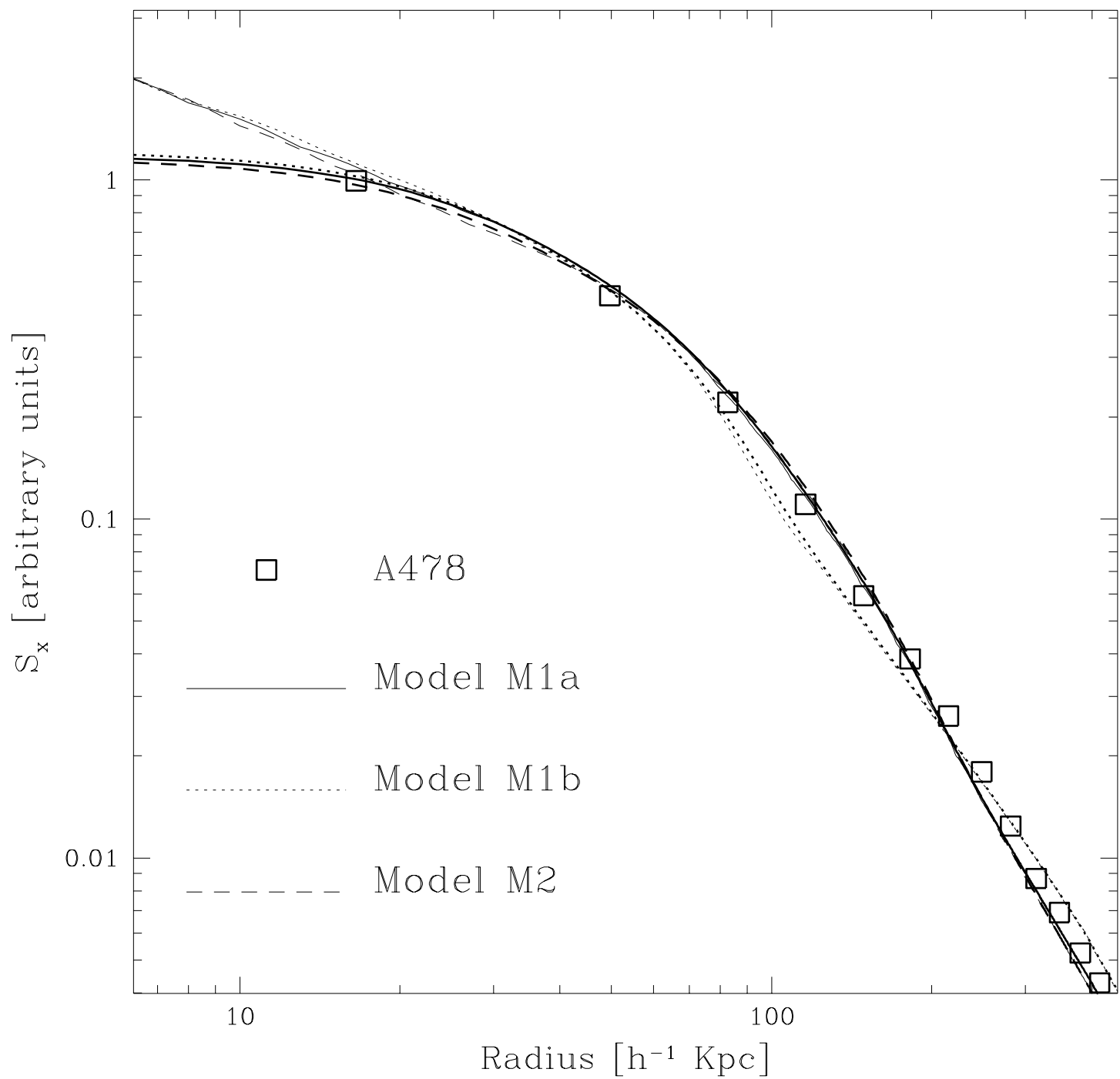

Fig. 4

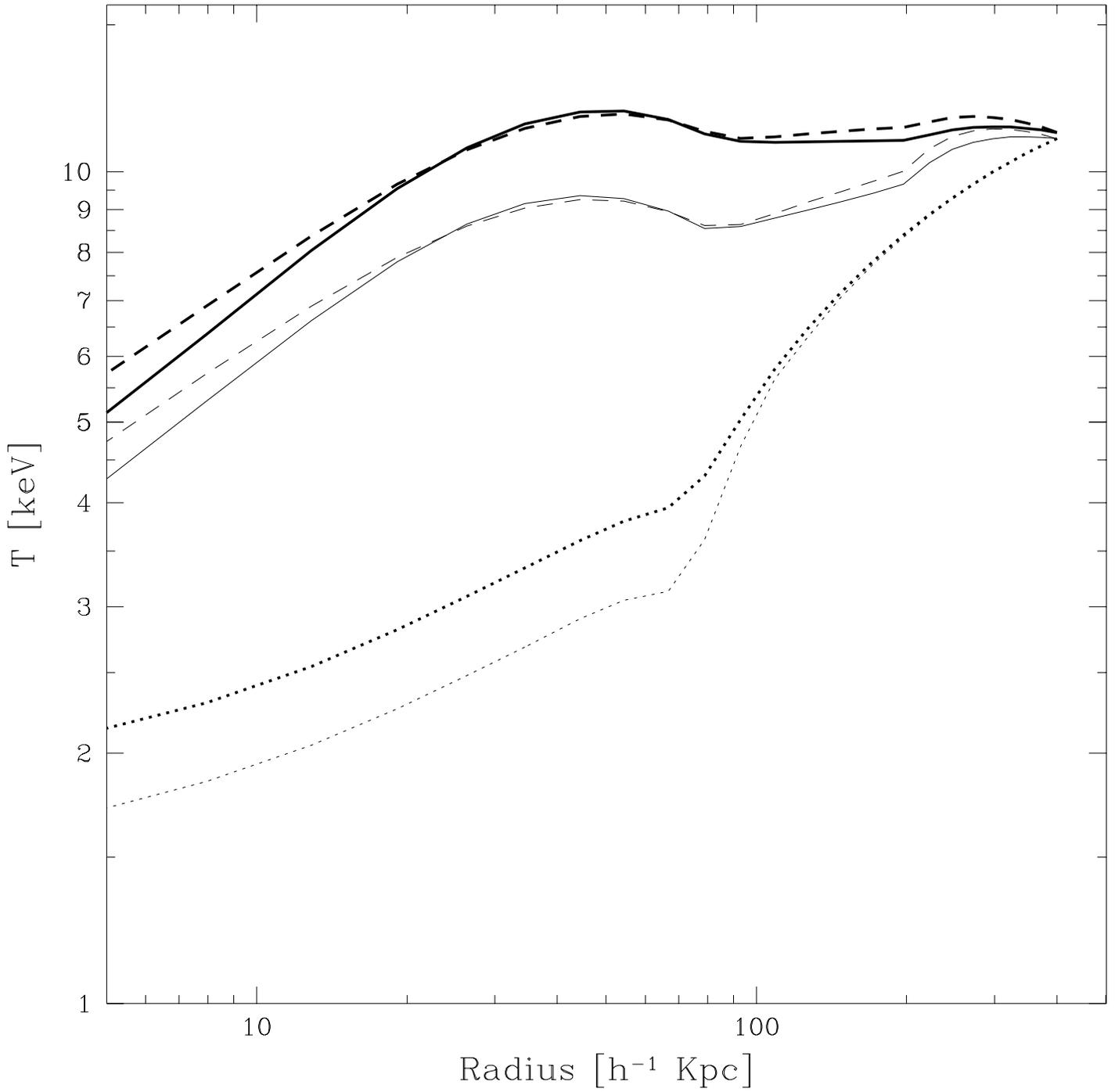

Fig. 5

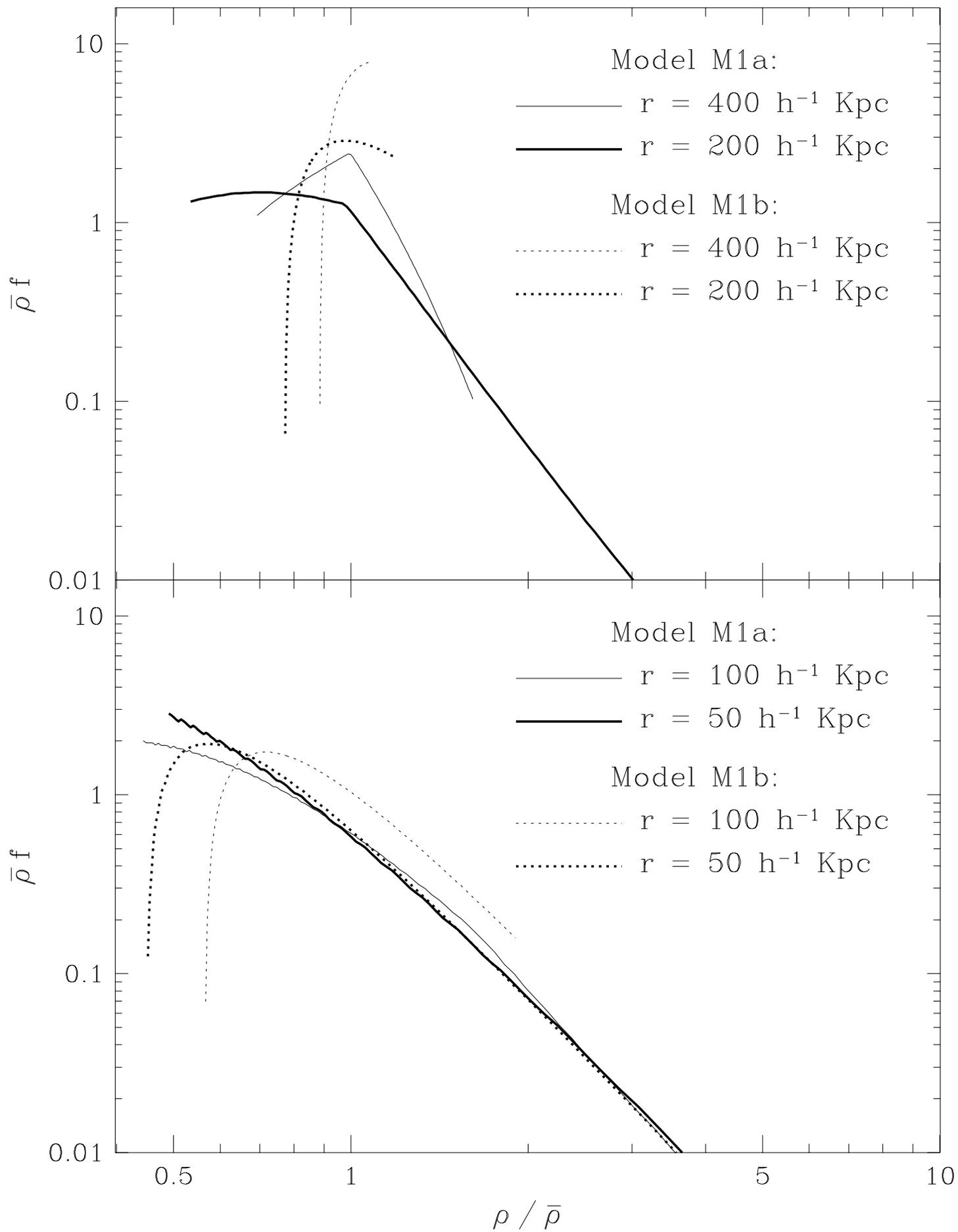

Fig. 6

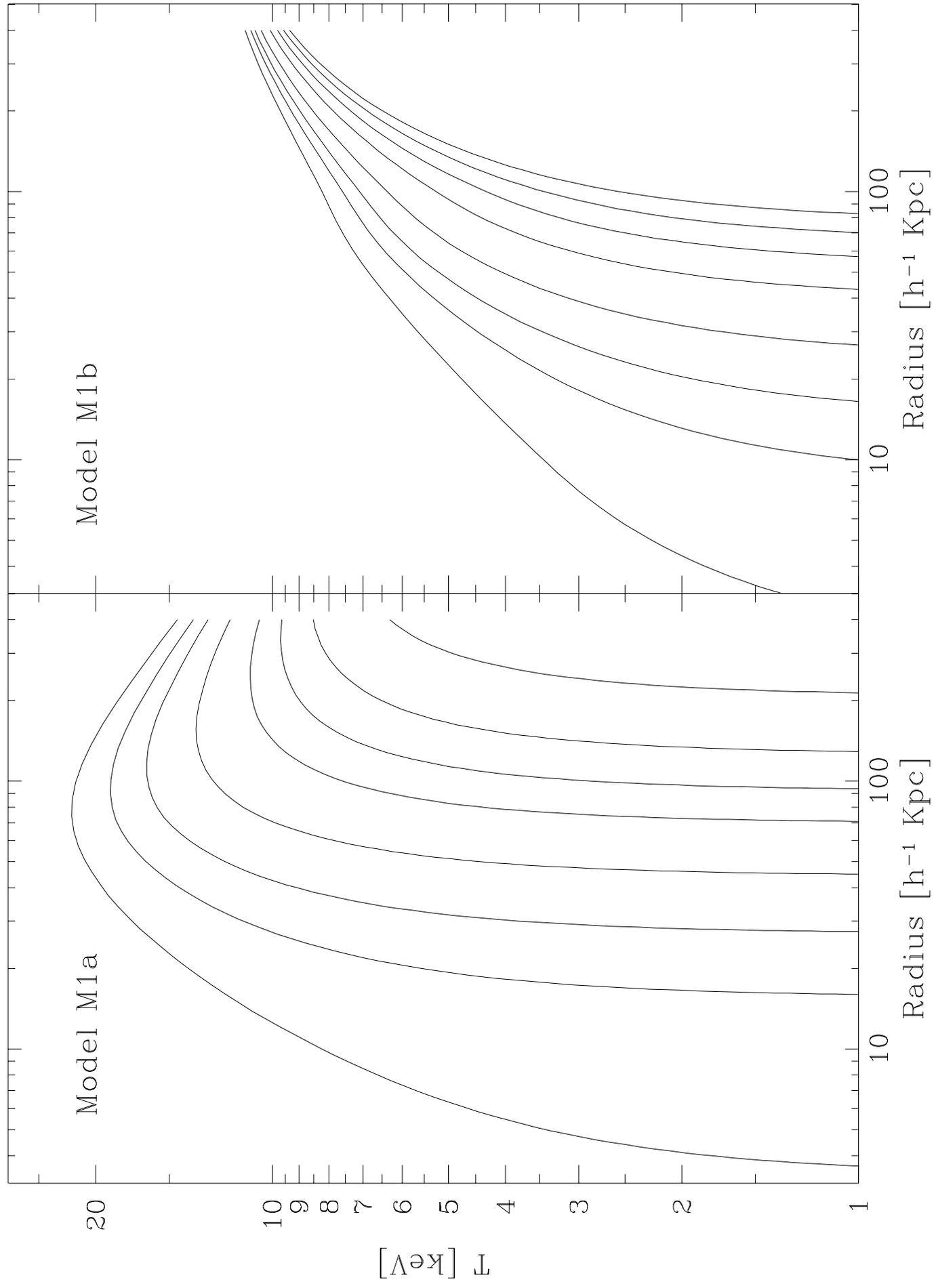

Fig. 7

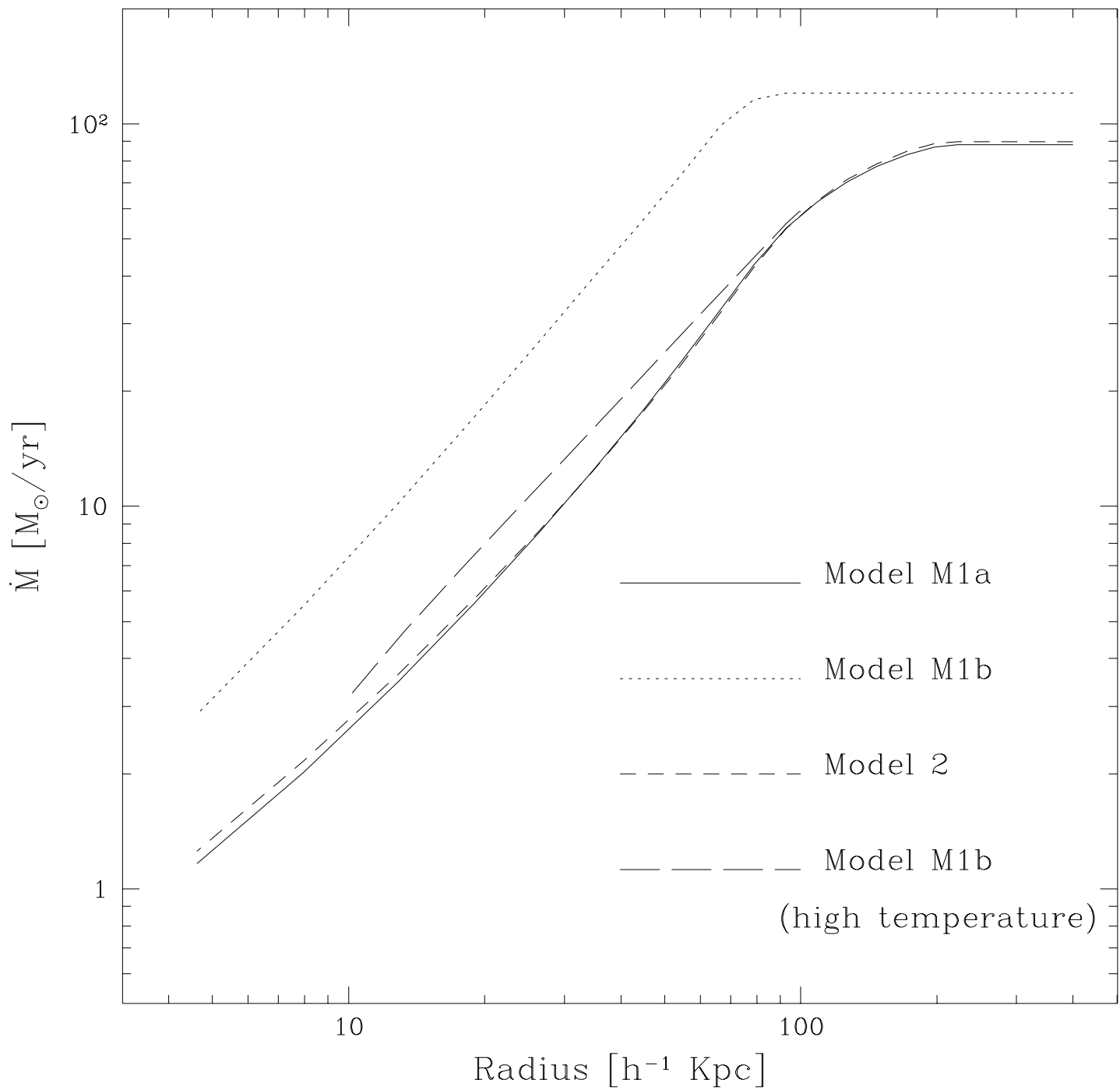
Fig. 8a

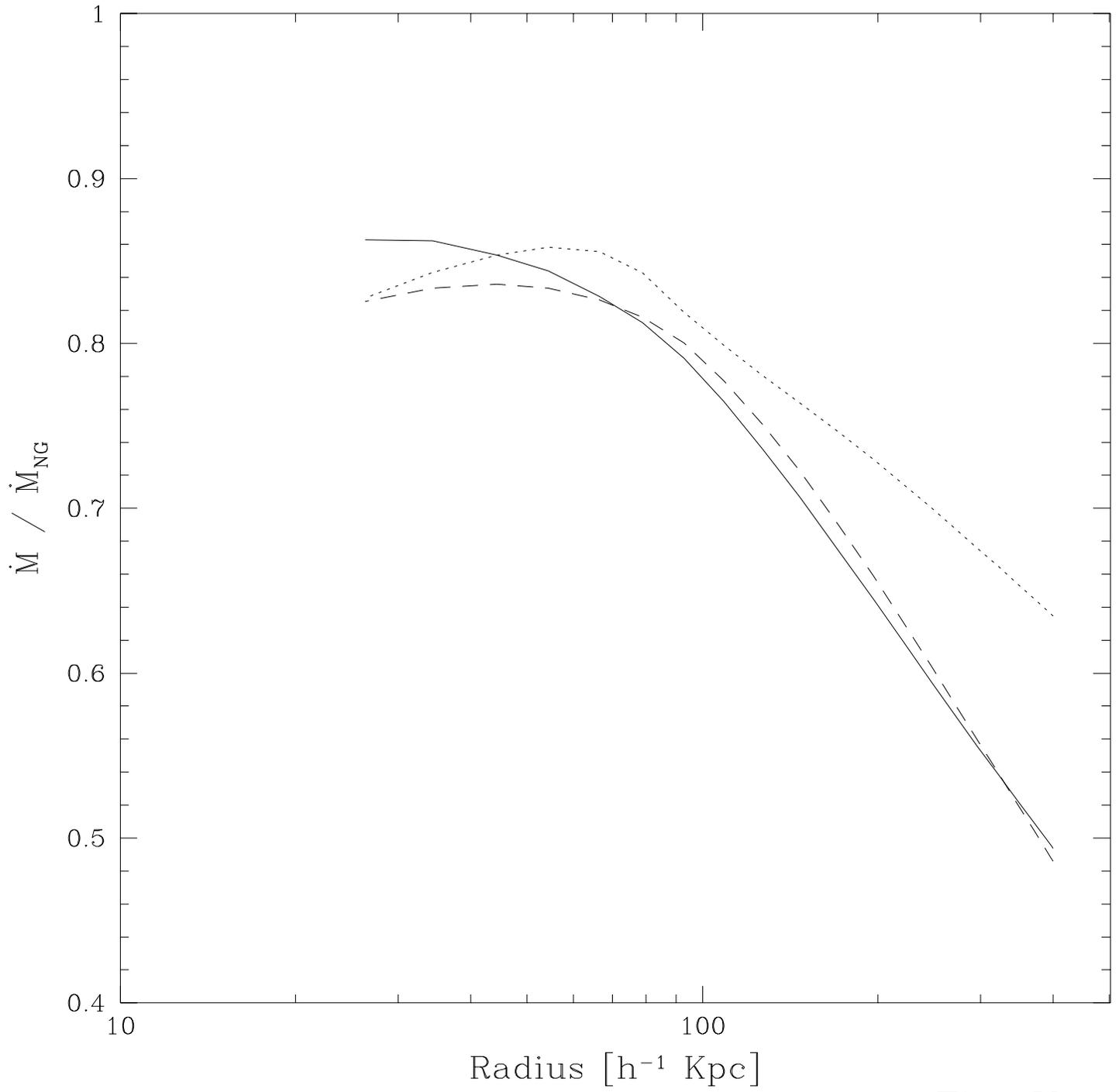

Fig. 8b

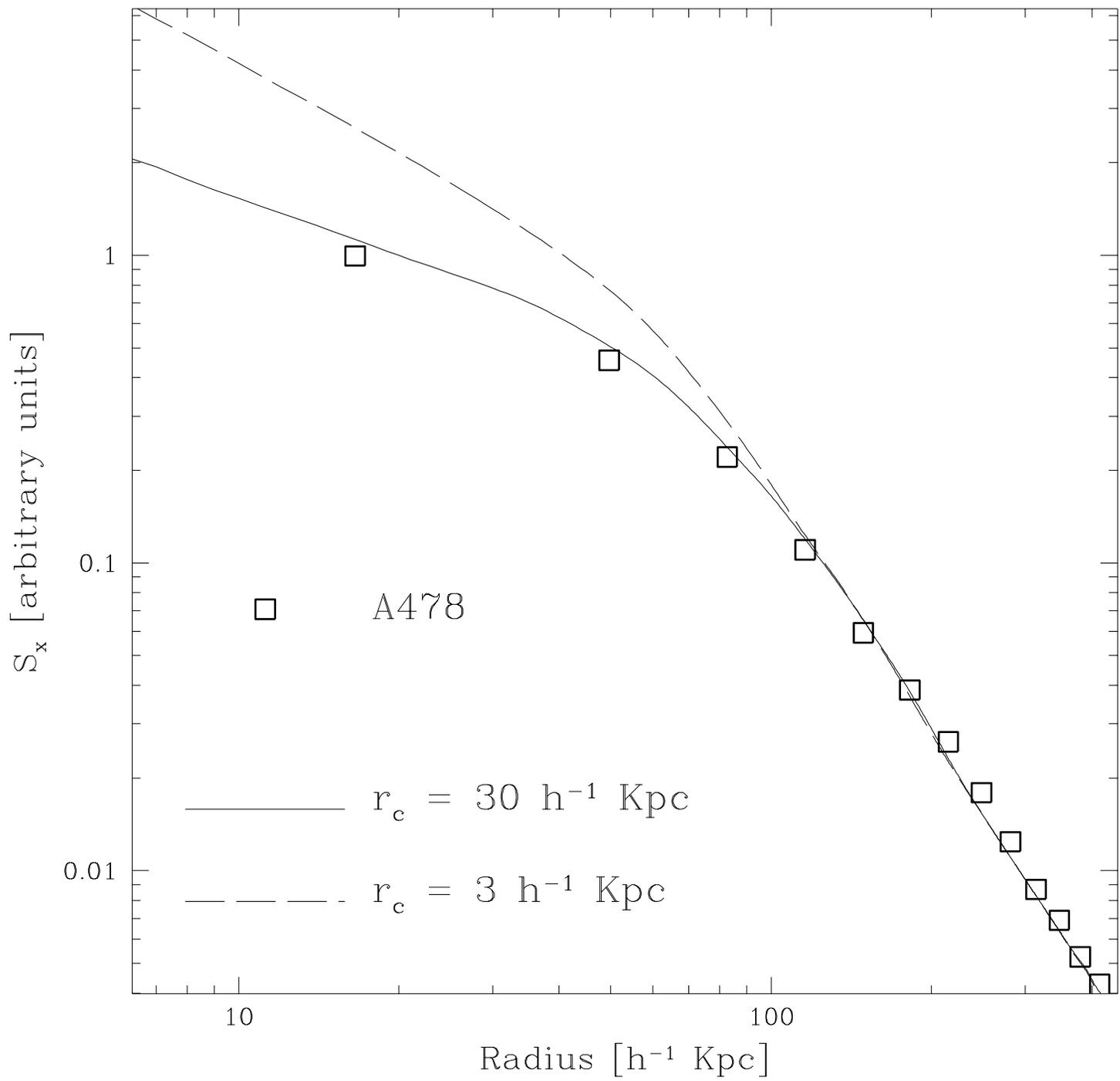

Fig. 9a

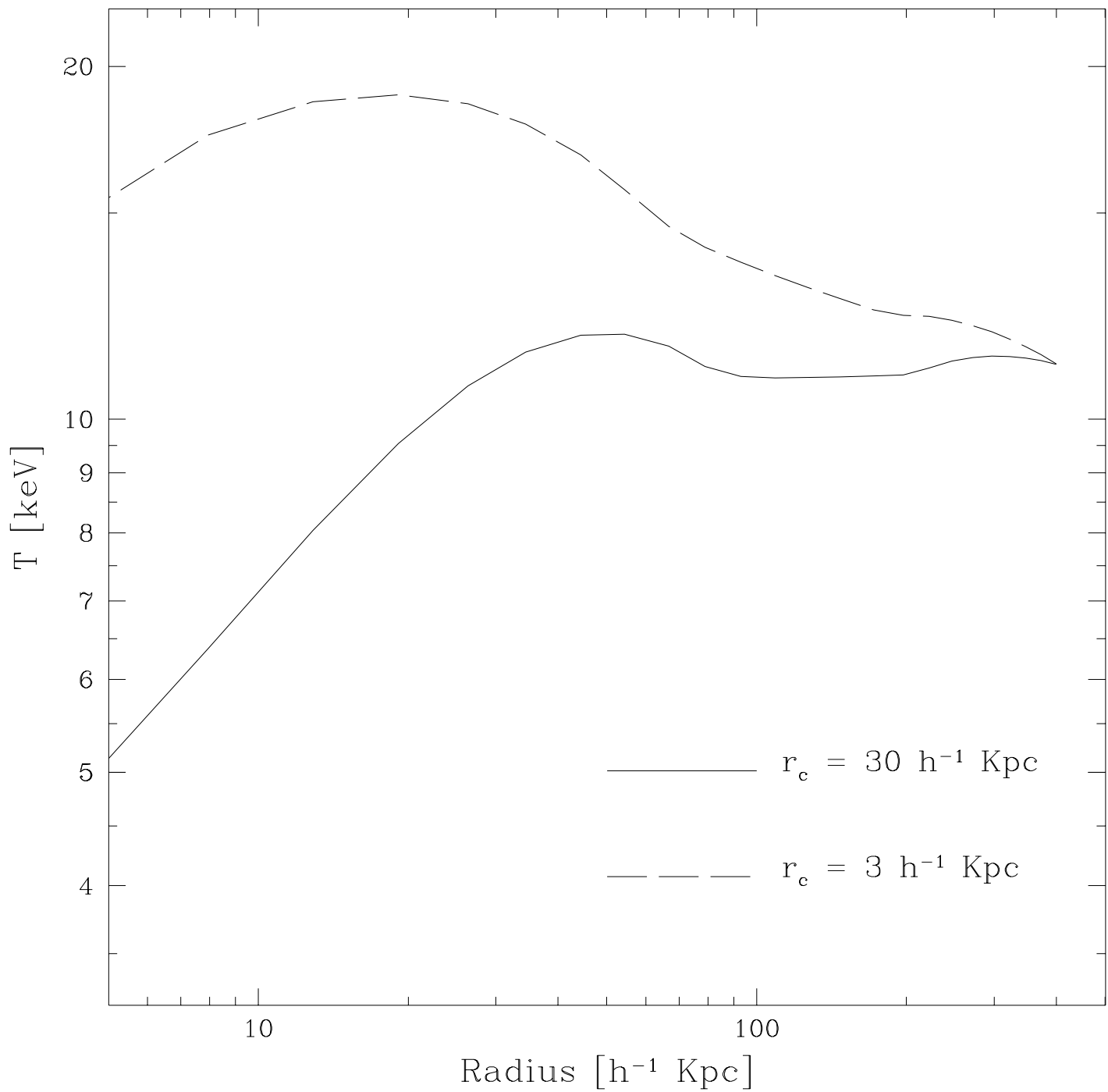

Fig. 9b

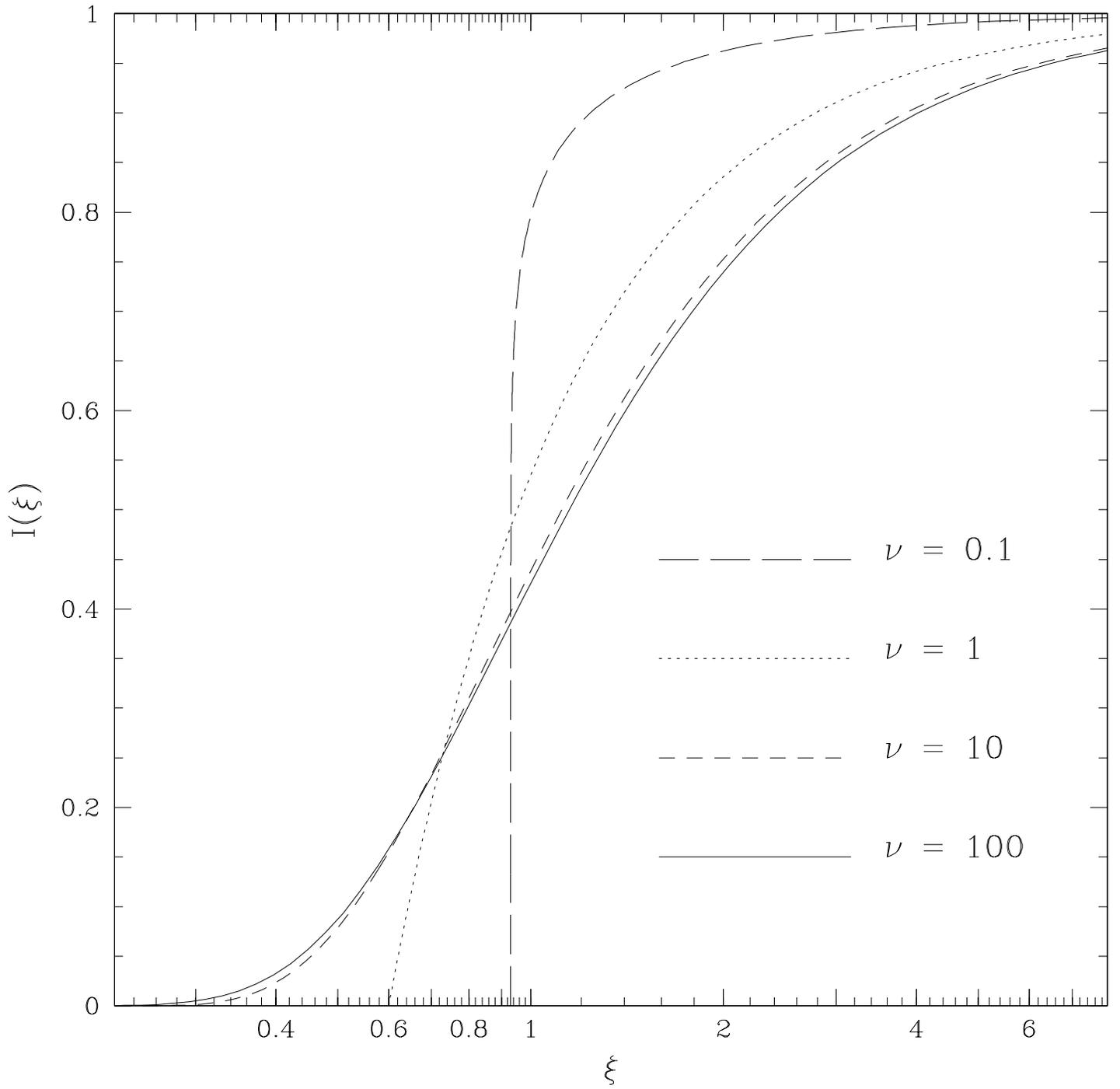

Fig. 10

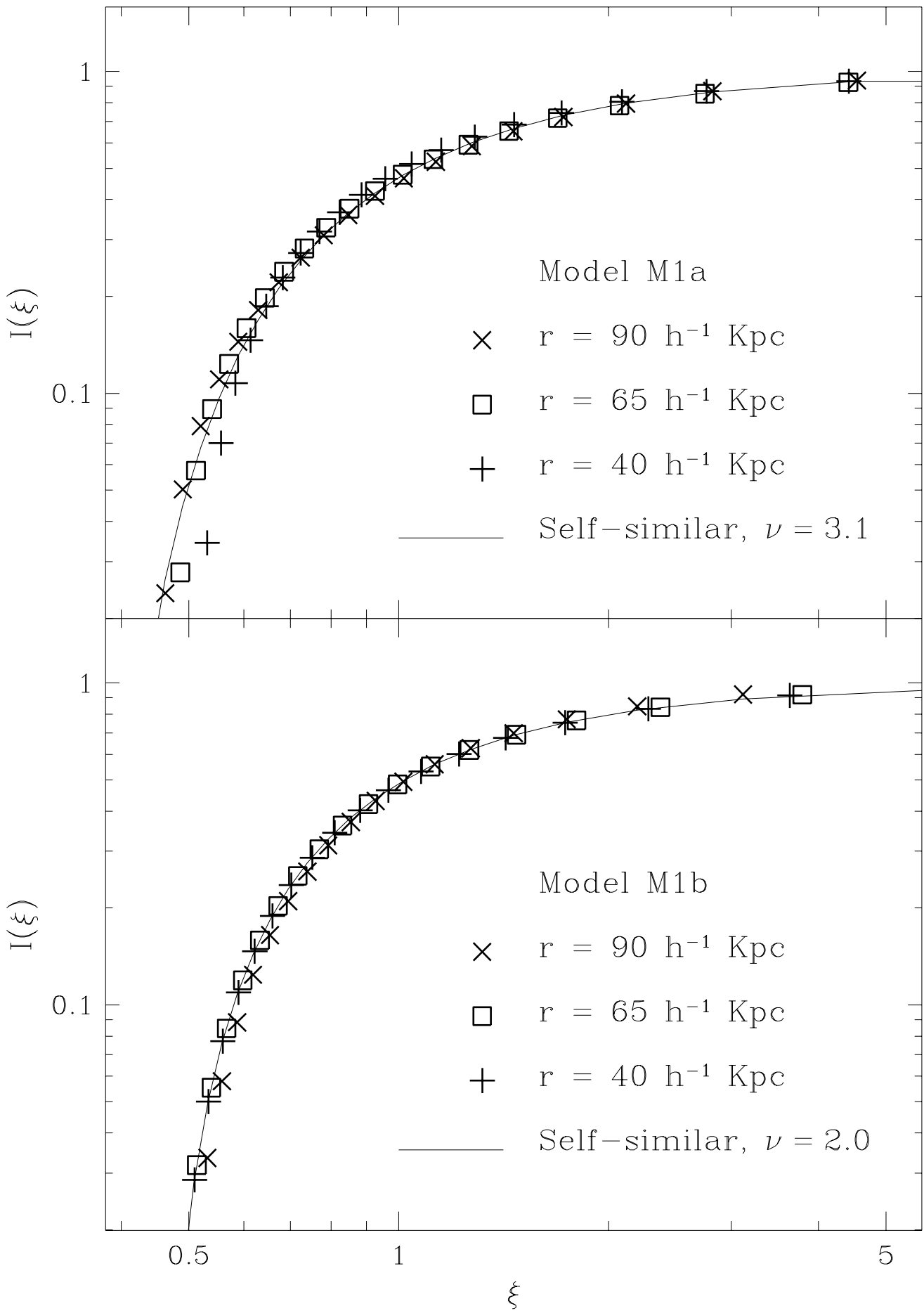

Fig. 11

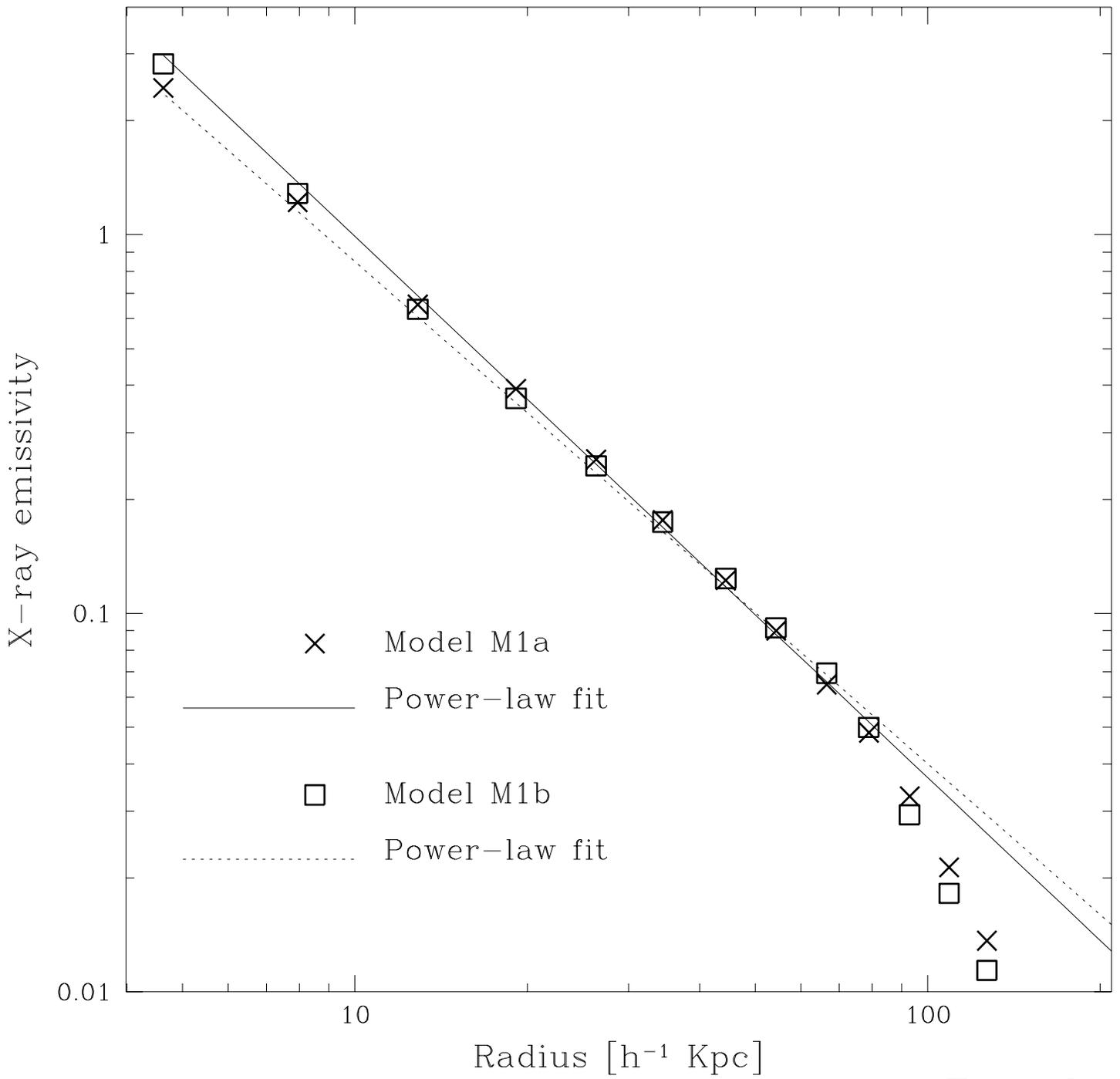

Fig. 12

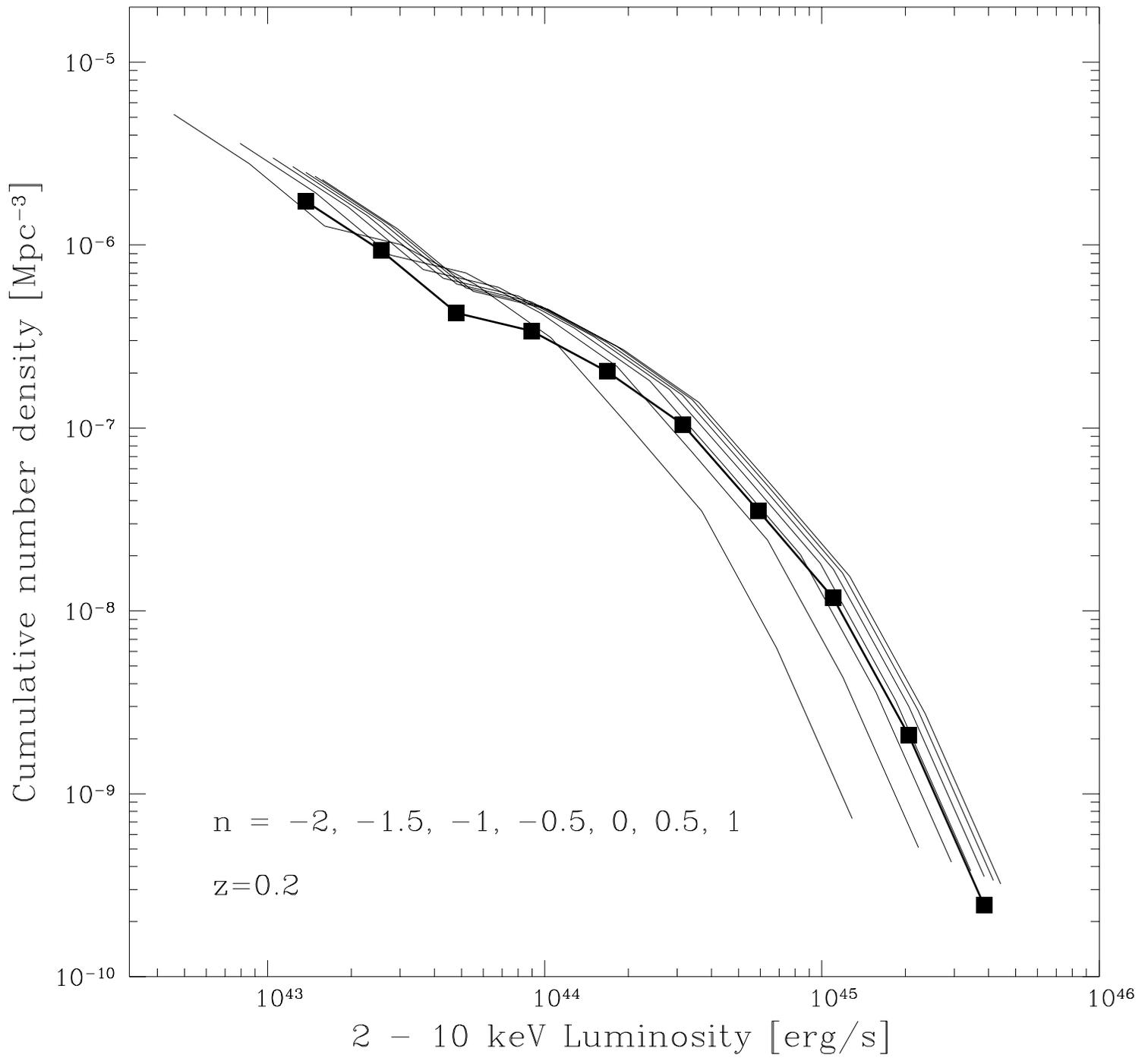

Fig. 13

# THE DISTRIBUTION OF MASS AND GAS
# IN THE CENTER OF CLUSTERS OF GALAXIES
# IMPLIED BY X-RAY AND LENSING OBSERVATIONS

Eli Waxman and Jordi Miralda-Escudé
Institute for Advanced Study, Princeton, NJ 08540

## ABSTRACT

Observations of gravitational lensing indicate that the mass distribution in clusters of galaxies (where most of the mass is dark matter) is highly peaked towards the center, while X-ray observations imply that the gas is more extended near the center. At the same time, the short cooling times for the gas and the observed X-ray spectra have demonstrated that the gas is cooling in these central regions; this fact has often led one to expect that the gas temperature should be lower near the center, and therefore the gas should be more concentrated than the dark matter.

We show that, for the mass profiles that are implied by gravitational lensing, the gas temperature must remain approximately constant within the cooling region in order to have consistency with the observed X-ray profiles. The gas temperature must also be similar to the galaxy velocity dispersion, since the profiles of the galaxies and the gas at large radius are similar. This is a consequence of hydrostatic equilibrium, and is independent of cooling flow models. We then find that a multi-phase cooling flow naturally produces an approximately constant temperature profile, and a more extended distribution for the gas compared to the mass. Cool phases are deposited at relatively large radius, while hot phases are adiabatically heated as they flow inwards and can keep the average temperature constant. Thus, cooling flows result in an *increase* of the central temperature, relative to a case where there is no cooling and the gas follows the mass distribution. The total mass deposition rates are determined primarily by the emissivity profile and the temperature at the cooling radius. They are also more sensitive to the assumed cluster age than in previous models, which assumed much more extended mass profiles.

The increased central temperatures caused by cooling flows give a characteristic core radius to the gas profiles, which is of order the cooling radius. This provides a natural explanation for the typical cores observed in X-ray clusters. It also changes the rate of cluster evolution expected in self-similar hierarchical models, since cooling occurred over a larger fraction of the gaseous halos in the past. We show that the inclusion of the effects of cooling flows predicts negative evolution for the high-luminosity clusters if $n \lesssim -0.5$, as is observed, but positive evolution for clusters of low luminosity. Some clusters are observed to have anomalously large core radii; we propose that these are in the process of merging and are not in dynamical equilibrium.

*Subject headings*: galaxies: clustering - cooling flows - X-rays: sources - gravitational lensing

# 1. INTRODUCTION

The X-ray surface brightness profiles of most clusters of galaxies are sharply peaked, often towards the brightest galaxy (usually a cD) at the cluster center. The cooling time of the central gas is usually found to be shorter than the Hubble time, in $\sim 70 - 90\%$ of X-ray clusters (Edge & Stewart 1991a; Edge, Stewart, & Fabian 1992). It is then inferred that the cooling of the gas should lead to a "cooling flow", an inflow of the gas required to maintain pressure equilibrium (see Fabian, Nulsen, & Canizares 1991, and Fabian 1994 for reviews), assuming that there is no significant energy injection that could balance cooling, and that the gas has indeed been cooling for long enough in a stationary potential well. X-ray spectra in the cooling regions of clusters give evidence for the existence of cold components, at temperatures well below the average X-ray temperature of the intracluster medium (ICM) (Canizares et al. 1979, 1982; Mushotzky et al. 1981; Lea et al. 1982; Fukazawa et al. 1994), and give gas cooling rates consistent with those obtained from X-ray imaging (Mushotzky & Szymkowiak 1988; Canizares et al. 1988).

Cluster X-ray profiles are generally modeled under the assumptions that the gas is isothermal and in hydrostatic equilibrium at radius larger than the cooling radius, typically of order $\sim 100 h^{-1}$ kpc (we use $h = H_0/100\,{\rm km\,s^{-1}\,Mpc^{-1}}$), and that at smaller radius a steady cooling flow forms (Fabian & Nulsen 1977; Fabian et al. 1981; Mushotzky et al. 1981; Nulsen et al. 1982; Canizares et al. 1982; Jones & Forman 1984; Stewart et al. 1984; White & Sarazin 1987a,b,c; Edge & Stewart 1991a; Edge, Stewart, & Fabian 1992; Allen et al. 1993). It has been customary to assume that the mass profile is given by a "$\beta$-model", with a core radius in the range $100 - 200 h^{-1}$ kpc, which is sometimes fitted to the outer part of the X-ray profile assuming a constant gas temperature (e.g., Abramopoulos & Ku 1983; Jones & Forman 1984), and in other cases it has simply been fixed to some assumed value in this range. It has then been found that the gas temperature should decrease towards the center, in order to reproduce an excess of central X-ray emission with respect to a model where the gas is isothermal at all radii; this has been interpreted as a consequence of the cooling flow. The core radii of cluster mass distributions obtained from the analysis of X-ray imaging are typically of order $\sim 150 h^{-1}$ kpc (Abramopoulos & Ku 1983; Jones & Forman 1984).

The discovery of gravitational lensing in clusters of galaxies (Soucail et al. 1987, 1988; Lynds & Petrosian 1989) gave us a direct probe to the distribution of dark matter in the central parts of clusters of galaxies. The presence of multiple and highly magnified images gives evidence that the mass profiles of clusters of galaxies must be centrally concentrated, ruling out the presence of flat cores larger than $30 - 50 h^{-1}$ kpc, depending on the cluster (Grossman & Narayan 1989; Miralda-Escudé 1993; Wu & Hammer 1993; Mellier et al. 1993; Kneib et al. 1993, 1994). These limits generally imply that the mass profiles of these clusters are steeper than the density profiles of the gas inferred from X-ray imaging data. It is not clear if all clusters have similarly concentrated mass profiles as in those where lensing is observed, but this does not seem to be contradicted by observations, and is also expected from N-body simulations of the formation of clusters (e.g., Navarro, Frenk, & White 1994).

In this paper, we shall investigate how cooling flow models are changed when a steep mass profile for clusters, consistent with the observations of gravitational lensing, is assumed. First of all, we show in § 2 that in order for the gas to be more extended than the mass, in a way consistent with the observed X-ray surface brightness profiles and the gravitational lenses, it is enough to have a temperature profile that is approximately constant over most of the cooling region. This is a pure consequence of hydrostatic equilibrium, and is totally independent of the presence or absence of cooling flows. In § 3 we analyze numerical as well as analytical models of cooling flows, and show that a constant temperature profile is not at all contradictory with the presence of a cooling flow. The general consequences of small core radii of the mass distribution for cooling flow models are described in detail. In § 4 we study the consequences implied for the evolution of X-ray clusters, and § 5 proposes a new physical interpretation for the classification of clusters. The main results are summarized in § 6.

## 2. DENSITY PROFILES FOR ISOTHERMAL GAS

In this Section, we shall see what the expected gas profiles should be if the gas is isothermal and in hydrostatic equilibrium in a potential well that can produce the lensed images observed in some rich clusters of galaxies.

Consider the spherically symmetric density profile for the cluster mass which has been most often used for gravitational lensing models:

$$\rho_M = \frac{\rho_{M,0} r_c^2}{r^2 + r_c^2} = \frac{\sigma^2/2\pi G}{r^2 + r_c^2} \ . \tag{1}$$

Here, $\rho_{M,0}$ is the central density, $r_c$ the core radius, and $\sigma$ is the velocity dispersion of the mass at $r \gg r_c$. We call this profile Model 1. As a typical case of lensing in clusters, we consider a lens at a redshift $z_l = 0.3$, with a critical radius $b = 15''$ for a source at redshift $z_s = 1$ (implying that the average surface density of the cluster within the radius $b$ is equal to the critical surface density for these redshifts). We choose a core radius $r_c = 30 h^{-1}$ kpc; for $\Omega = 1$, this corresponds to an angular size of $11''$. This is close to the maximum core radius consistent with the presence of arcs at $b = 15''$, since a flat core larger than $b$ would cause the arcs to be thick and with low curvatures, and imply an unrealistically large mass for the cluster (Grossman & Narayan 1988; Hammer 1991; Miralda-Escudé 1993). For this value of $r_c$, a velocity dispersion $\sigma = 1325$ km s$^{-1}$ in eq. (1) is required in order that the average surface density within the radius $b$ be equal to the critical value.

This example is similar to the observed values in the cluster MS 2137-23 (Fort et al. 1992; Mellier, Fort, & Kneib 1993), although the redshifts of the arcs in this cluster have not yet been determined. In this particular cluster, the critical radius is not only determined from the position of a long arc, but also from five different images of two background sources. This makes the lensing model for this cluster very robust, and confirms the value for the critical radius. At the same time, the presence of the radial arc constrains the value of the core radius to be not much smaller than the value we assume here; this is also required in order that the stellar velocity dispersion of the central cD galaxy can be reasonably small (Miralda-Escudé 1994).

Let us now consider what density profiles for the gas are consistent with this small core radius, if the gas is isothermal and in hydrostatic equilibrium. For a given value of $r_c$, the gas profile is determined by the ratio $\beta \equiv \mu \sigma^2 / T$, where T is the gas temperature and $\mu$ is the average particle mass. In Figure 1a, we show the mass and gas density profiles for isothermal gas with the three indicated values of $\beta$. The lowest temperature (highest $\beta$) give, of course, the steepest profile (the normalizations of the profiles in this figure are arbitrary). For the two cases with higher temperature, the gas profile at large radius has a slope close to isothermal, consistent with the observed X-ray profiles in clusters of galaxies, whereas the case with $\beta = 1.7$ is too steep. At small radius, $r \sim 100h^{-1}$ kpc, the gas profile flattens faster than the mass profile.

The fact that the gas profiles of these isothermal models are more extended than the mass profile near the center is because, for a mass density profile which flattens continuously toward the center from an asymptotically isothermal slope at large radius, the velocity dispersion of the mass must decrease towards the center. If the gas is isothermal, it must therefore be hotter than the mass at the center, as long as the gas follows the mass at large radius (notice that the solution of a self-gravitating isothermal sphere with a finite flat core requires the slope of the density profile to be steeper than $r^{-2}$ in a region just outside the core, in order to maintain a constant velocity dispersion; see Figs. 4.7 and 4.8 in Binney & Tremaine 1987). Thus, having a gas distribution which is more extended than the mass, as required in clusters where lensing is observed, does not require a substantial increase of the gas temperature towards the center, but is instead consistent with a temperature that is approximately constant down to small radius.

We now consider a second model for the density profile, which we call Model 2:

$$\rho_M(r) = \frac{\rho_{M,0}\, r_c^3}{r\,(r + r_c)^2} \ . \tag{2}$$

This profile has been found to fit well the density profiles of simulated clusters of galaxies in the Cold Dark Matter model with $\Omega = 1$, for clusters which have not experienced a recent merger (Navarro et al. 1994). It is also found in these simulations that the density $\rho_{M,0}$ is approximately constant between different clusters (namely, Navarro et al. find $\rho_{M,0} = 7500\, \bar{\rho}$, where $\bar{\rho}$ is the critical density; their radius $r_{200}$ is $5r_c$), so that there is only one parameter for the profiles of clusters of galaxies. If we calculate the surface density for this model, we find that in order to have a critical radius of $b = 15''$ in the same example of a cluster lens used above (i.e., for a cluster redshift $z_l = 0.3$ and a source redshift $z_s = 1$, and using also the value of $\rho_{M,0}$ at $z = 0.3$), we need $r_c = 265h^{-1}$ kpc (in proper units; this will be done more generally in Miralda-Escudé & Navarro 1995, in preparation). This corresponds to a mass velocity dispersion of $\sigma = 1305$ km s$^{-1}$, as found by scaling the velocity dispersions given in Table 1 of Navarro et al. to this core radius and redshift, which is almost the same as for our previous model. The density profile of this model, which we call Model 2, is given in Figure 1b, together with the gas profiles at the indicated $\beta$ values. The conclusions are similar as for Model 1. In Model 2, the value of $r_c$ is much larger than in Model 1, but the $r^{-1}$ cusp in the density profile allows for a sufficient central mass to account for lensing. The different shape of Model 2 with respect to Model 1 does not affect very much the form of the gas density profile. Basically, the core of the isothermal gas will be at the radius where the mass velocity dispersion starts decreasing to the center, and this occurs at a similar radius in both models when they are required to have the same critical radius for lensing, and their average mass velocity dispersion is similar.

The X-ray profiles given by these gas density profiles are obtained by projecting the emissivity, which is proportional to the square of the gas density. These are shown in Figure 2, for the same models as in Figure 1 but only for $\beta = 1$. The open squares give the X-ray profile for the cluster A478 (at z=0.088) observed with the ROSAT PSPC, which we have taken from Allen et al. (1993). We also show, as the solid line, the convolution of the X-ray profile for Model 2 with a gaussian with FWHM $= 33 h^{-1}$ kpc, corresponding to $30''$ which is approximately the PSPC point spread function. A478 is an example of a "cooling flow" cluster; its cooling time near the center is less than the Hubble time, and the X-ray spectrum gives evidence for the presence of cool gas (Allen et al. 1993). In general, other clusters classified as having cooling flows have very similar X-ray profiles, with core radii in the range $r_c = 100 - 200 \, h^{-1}$ kpc (Ku et al. 1983; Jones & Forman 1984; Stewart et al. 1984; Edge & Stewart 1991a). The centrally peaked X-ray emissivity has often been thought to arise as a result of the cooling flow. However, we see that with a more centrally concentrated mass distribution, which is in agreement with the constraints on the mass profiles in lensing clusters, the observed X-ray profiles can be reproduced when the temperature is constant, for $\beta \simeq 1$.

Although the X-ray surface brightness profile of our examples for lensing clusters with isothermal gas are similar to the observed ones, the temperatures required to have the observed critical radii are very high. For $\beta = 1$, the X-ray temperature of the gas corresponding to the velocity dispersions of our models, which give a critical radius $b = 15''$ for a cluster at $z = 0.3$ and a source at $z = 1$, is 11 keV. Very few clusters are known having such high temperatures. For example, A478 has a temperature of 6.5 keV (Allen et al. 1993). We can solve this problem by simply dividing the density of the cluster models by a constant factor at all radii; if the temperature is divided by the same factor, the gas density and X-ray profiles in Figures 1 and 2 are not altered. However, if the cluster is less massive, then the critical radius for lensing is reduced. The lensing cluster MS2137-23 has the critical radius assumed in our example, but it is *less* luminous than A478, and its temperature is unlikely to be higher.

Such a difficulty in reconciling the X-ray temperatures with the mass profiles inferred from lensing was encountered in the analysis of the clusters A2218, A1689 and A2163 of Miralda-Escudé & Babul (1994, 1995). It was found that in order for the gas to be in hydrostatic equilibrium in a potential well consistent with the observed lenses, the temperature needs to be about two times larger than observed for the first two clusters, while it is consistent with the observed value for the third one. As we have shown here, the root of this problem is that the X-ray temperatures are too low; the fact that the gas is more extended than the mass is, by itself, not a problem, as long as the temperature can be constant within the gas core. As discussed in Miralda-Escudé & Babul, there are several effects that can explain these low temperatures. First, it may be that the mass estimated from gravitational lensing is too large. This could arise from projection effects, possibly combined with substructure, when the cluster is elongated along the line of sight. Substructure also implies that bulk motions are present in the gas (which give additional support against the potential well), and that the gas is not in hydrostatic equilibrium. In fact, in A2218, the HRI ROSAT image suggests the presence of central substructure (Kneib et al. 1994). These effects alone can probably explain the discrepancy between the lensing masses and the X-ray temperatures, and in this case the masses derived from X-ray temperatures and hydrostatic equilibrium models in most relaxed, X-ray selected clusters would be correct. Two other effects were also suggested in Miralda-Escudé & Babul, namely a systematic underestimate of the temperatures caused by a multiphase medium, and a contribution to the pressure from magnetic fields (see also Loeb & Mao 1994); if these are important, they could cause a systematic error in the derived masses even in relaxed clusters. In any case, here we shall assume that in relaxed clusters the central gas is indeed in hydrostatic equilibrium, and that magnetic fields are dynamically unimportant.

A large fraction of X-ray clusters have a central cooling time of the gas shorter than the age of the cluster, and are classified as containing cooling flows (Edge et al. 1992). It might seem that this should cause the temperature within the cooling region to decline toward the center, and therefore the X-ray profiles to be much more peaked than in an isothermal model. In fact, we shall see in the rest of the paper that cooling flow models predict the temperatures to be approximately constant over most of the region where the gas is cooling, for mass profiles with core radii that are consistent with lensing observations, and that as a consequence the distribution of gas in this region should be, as observed, more extended than that of the mass.

## 3. COOLING FLOWS

### 3.1. *Multiphase Cooling Flow Model*

We consider a spherically symmetric steady state cooling flow. As we mentioned in § 1, X-ray spectra of the brightest emission regions in clusters imply that the intracluster medium (ICM) contains gas over a range of temperatures well below the average temperature of the X-ray emission, at a range of radii (e.g., Fabian, Nulsen, & Canizares 1991). This, in turn, implies that the ICM is inhomogeneous and should be described at any given radius as a multiphase medium consisting of gas blobs of various densities. We define a volume distribution function $f(r, \rho)$, so that at a radius $r$ the volume fraction occupied by gas in the density range $\rho$ to $\rho + d\rho$ is $f(r, \rho)d\rho$.

In constructing the equations describing the multiphase flow, an important simplification that can be made is that the flow velocity $u$ is typically much smaller than the speed of sound. This implies that the inertial term in the momentum equation can be neglected, and also that the various gas phases are approximately in pressure equilibrium. The pressure $p$ may therefore be assumed to be a function of $r$ alone. This assumption only breaks down for very dense blobs, which have cooled below X-ray emitting temperatures and occupy a small fraction of the volume.

A second simplification that was proposed by Nulsen (1986, hereafter N86) and Thomas (1988), consists in assuming that all the phases with different temperatures are effectively comoving with the mean flow, and they move inwards with the same radial velocity $u(r)$, which depends only on the radius. Nulsen (1986) argued that dense, large blobs which move fast relative to the ambient flow due to their own weight should be quickly disrupted owing to the Kelvin-Helmholtz instability, until the blobs are so small that friction would make their motion negligible. In addition, the cool phases need to be insulated by magnetic fields in order to prevent their evaporation and the consequent destruction of the multiphase medium, and these magnetic fields can also prevent the motion of the dense blobs relative to the lighter phases. Magnetic fields are also needed to make the cooling gas thermally unstalbe in the linear regime (Balbus & Soker 1989; Loewenstein 1990; Balbus 1991). However, it is not clear if the blobs should still fall at a significant terminal velocity, depending on the occurrence of magnetic field reconnection. At the same time, in order for the multiphase medium to be stable, there should be a typical size of the blobs for which disruptive processes are balanced by growth mechanisms, such as merging and accretion of the surrounding gas due to a highly reduced thermal conduction in the presence of magnetic fields. These possible complications will be ignored in this paper, since there is at present no predictive theory to describe them.

As the gas flows inward the density and entropy of each phase change. We denote by $\rho(r, r_0, \rho_0)$ the density at radius $r$ of the phase which had a density $\rho_0$ at some outer radius $r_0$, and by $S(r, r_0, \rho_0)$ the specific entropy of this phase at radius $r$. The radial dependence of $\rho$ and $S$ is governed by the momentum equation, which reduces with the above assumptions to the hydrostatic equilibrium equation

$$\frac{dp}{dr} = -\bar{\rho}(r)\frac{GM(r)}{r^2}, \tag{3}$$

and by the energy equation

$$\rho u T \frac{dS}{dr} = -\rho^2 \Lambda(T). \tag{4}$$

Here, $\bar{\rho}(r)$ is the volume averaged density at radius $r$, $M(r)$ is the gravitating mass enclosed within $r$, $T$ is the temperature of the phase, and $\Lambda(T)$ is the collisional cooling function, defined so that $\rho^2 \Lambda$ is the energy lost per unit time and per unit volume.

Following N86 we define the mass flux function $\psi$ as

$$\psi(r, \rho) = 4\pi r^2 u \int_0^\rho \rho' f(r, \rho') \, d\rho'. \tag{5}$$

The function $\psi(r, \rho)$ gives the mass flow rate at radius $r$ at densities up to $\rho$. The total accretion rate is $\dot{M}(r) = -\psi(r, \infty)$ (note that for inflowing gas $u$ and therefore $\psi$ are negative while $\dot{M}$ is positive). Conservation of mass implies that the rate of mass inflow at densities up to that of a certain phase is independent of $r$, i.e., that

$$\frac{\partial \psi}{\partial r} + \frac{d\rho}{dr}\frac{\partial \psi}{\partial \rho} = 0. \tag{6}$$

The velocity $u$ and average density $\bar{\rho}$ are determined from the mass flux function by

$$u(r) = \int_0^\infty \frac{1}{4\pi r^2 \rho}\frac{\partial \psi}{\partial \rho} d\rho, \tag{7}$$

and by

$$\bar{\rho}(r) = \psi(r, \infty)/[4\pi r^2 u]. \tag{8}$$

Thus, the multiphase flow solution is completely described by $\psi(r, \rho)$ and $p(r)$. Equations (3)-(4) and (6) determine the radial evolution of $\psi$ and $p$. These equations must be supplemented with a cooling function $\Lambda(T)$.

The main objective of this work is to analyze the dependence of the cooling flow structure on the mass profile $M(r)$, and specifically on the assumed core radius of $M(r)$. For this purpose it is sufficient to use the simple approximation

$$\Lambda(T) = \Lambda_0 T^\alpha \tag{9}$$

for the collisional cooling function. This simple expression gives a reasonable approximation to the cooling function for X-ray temperatures while allowing the derivation of general analytic results.

The following point should be noted when solving the flow equations. Using (9) down to $T = 0$ causes the density of any phase to diverge at some finite radius. Thus, phases whose density diverge should be continuously "removed" from the flow. The removed phases have usually been referred to as "deposited" phases in the literature. Such a "removal", or "deposition", is done by defining $\rho_c(r, r_0)$ as the density at $r_0$ of the phase which reaches infinite density at the smaller radius $r$. The definition (7) of $u$ in terms of $\psi$ should then be replaced with

$$u(r) = \int_0^{\rho_c(r,r_0)} \frac{1}{4\pi r^2 \rho(r,r_0,\rho_0)} \frac{\partial \psi[r, \rho(r,r_0,\rho_0)]}{\partial \rho_0} d\rho_0, \qquad (10)$$

and $\psi(r, \infty)$ in (8) is replaced with

$$-\dot{M}(r) \equiv \psi(r, \infty) = \psi[r_0, \rho_c(r, r_0)]. \qquad (11)$$

With this definition $\dot{M}$ is the mass flow rate of gas that has not yet been deposited.

The form of the flow equations allows to construct a family of cooling flow solutions by scaling the variables of a given single solution. Such scalings conserve the *shapes* of the emissivity profile $\epsilon(r)$ and cluster mass profile $M(r)$, while allowing to change the luminosity, temperature, and physical length scale of the solution independently. Scaling the flow variables $u$, $\rho$, and $T$, the mass profile $M$ and the length scale of a given solution gives a new solution,

$$\rho'(r) = \chi_\rho \rho(r/\chi_r), \qquad u'(r) = \chi_u u(r/\chi_r), \qquad T'(r) = \chi_T T(r/\chi_r), \qquad (12)$$

for

$$M'(r) = \chi_M M(r/\chi_r), \qquad (13)$$

provided that the mass and velocity scalings are related to those of the other variables as

$$\chi_M = \chi_T \chi_r, \qquad \chi_u = \chi_T^{\alpha-1} \chi_\rho \chi_r. \qquad (14)$$

Thus, any solution of the cooling flow equations can be used as a model for a cluster with any luminosity, temperature and core radius of the X-ray surface brightness by using these scalings, without changing the form of the solution. This property will be used in the rest of this section.

### 3.2. Numerical Solutions

We have numerically solved the multiphase flow equations for three representative mass profiles. Two of them are the mass models which were described in §2 and are consistent with the presence of gravitational lensing. The models M1a and M2 are the same as the two models used in § 2: M1a has the form (1) with $\sigma = 1325 \, \text{km s}^{-1}$ and $r_c = 30 h^{-1}$ kpc, and M2 has the form (2) with $r_c = 265 h^{-1}$ kpc and $\sigma = 1305 \, \text{km s}^{-1}$. The third model, denoted M1b, has the same form and velocity dispersion of M1a, but a larger core radius $r_c = 125 h^{-1}$ kpc. Model M1b is representative of those that have been normally used for constructing cooling flow models for X-ray clusters (e.g., Edge & Stewart 1991a). All models include a central galaxy characterized by a de Vaucouleurs law (tabulated by Young 1976), an effective radius of $20 h^{-1}$ kpc and a mass of $1.7 \cdot 10^{12} \, M_\odot$. This is, again, typical of cooling flow models in the literature (Wise & Sarazin 1993; White & Sarazin 1987c). The mass profiles of the three models are compared in Figure 3. We notice that Models M1a and M2 have a very similar mass profile, and therefore their cooling flow solutions will also be similar. Note that using the scaling laws (12-14) the solutions presented below are easily scaled to any $H_0$ value. Such a scaling would not affect the radial shape of the various flow variables, nor the temperature scale, nor the lensing properties; however, it affects the position of the cooling radius.

We have integrated the equations of the steady state cooling flow starting at some external radius $r_0$. The solution inside this radius is determined by the pressure $p_0$, the density distribution $f_0(\rho)$, and the mass flow rate $\dot{M}_0$ at $r = r_0$. Given an outer density distribution $f_0(\rho)$, we fix the outer pressure $p_0$ by requiring that the emission averaged temperature at $r_0$ be equal to the cluster virial temperature of 11 keV (given the velocity dispersion of the mass models). This ensures that, outside $r_0$, the gas will have an approximately isothermal slope if the temperature is constant, which is generally observed in clusters. The outer mass flow rate $\dot{M}_0$ is determined by the condition that $\dot{M}(r) = 0$ at $r = 0$. An initial value $\dot{M}_0$ is chosen, and the solution is iterated until the requirement that the mass flow rate is zero at the center is satisfied.

The outer density distribution $f_0(\rho)$ determines the form of the cooling flow solution and, in particular, the X-ray emissivity profile $\epsilon(r)$. Specifically, the *shape* of the outer distribution $f_0(\rho)$ determines the *shape* of the emissivity profile $\epsilon(r)$, while the average density $\bar{\rho}_0$ determines the normalization of $\epsilon$, i.e., the total luminosity. This is easily seen from the scalings (12-14). We have numerically searched for density distributions $f_0(\rho)$ for the three mass models, that give an emissivity $\epsilon(r)$ consistent with observed surface brightness profiles in typical clusters with short central cooling times. The outer radius $r_0$ was chosen for the numerical integrations to be $r_0 = 400 h^{-1}$ kpc. For the cooling function we have used $\alpha = 0.5$ and $\Lambda_0 = 2.4 \cdot 10^{21}\,\mathrm{erg\,s^{-1}\,cm^{-3}\,g^{-2}\,K^{-0.5}}$. The gas equation of state was taken to be that of a perfect gas with average mass per particle of $\mu = 10^{-24}$ g. These values correspond to ionized plasma with the cosmic helium abundance, and at high enough temperatures to make the free-free cooling dominant over line emission by metal species.

The calculated surface brightness profiles are plotted in Figure 4 for the three mass models. We have not included any temperature dependence of the emissivity to calculate the X-ray surface brightness, since for typical cluster temperatures and the sensitivities of most X-ray instruments the detected X-ray counts depends very little on the temperature of the gas; we have simply assumed the emissivity to be proportional to the square of the gas density. These are compared with the same X-ray profile of A478 (Allen et al. 1993) as in Fig. 2; the models are reasonably close to this observed cluster. The X-ray temperature of A478 is 6.5 keV, lower than the 11 keV assumed in our models. As described at the end of §3.1, the temperature of the solution may be scaled without affecting the surface brightness profile, if the total mass of the cluster is scaled by the same amount at all radii (although of course, a cluster with lower mass is then not able to produce gravitationally lensed images, with the large deflection angles typically measured, as discussed in § 2).

We now describe the important qualitative characteristics of the solutions. The emissivity averaged temperature,

$$T_\epsilon(r) \equiv \int_0^\infty f(r,\rho) \rho^2 \Lambda(T) T \mathrm{d}\rho / \epsilon(r) \;,$$

and the mass averaged temperature,

$$T_M(r) \equiv \int_0^\infty f(r,\rho) \rho T \mathrm{d}\rho / \bar{\rho}(r) \equiv \frac{\mu p(r)}{\bar{\rho}(r)} \;,$$

are plotted in Figure 5. $T_\epsilon$ is lower than $T_M$ since it has higher contribution from the high density phases which cool rapidly. The gas temperature obtained for the models M1a and M2 is approximately constant for $r > 20h^{-1}$ kpc and declines significantly only at smaller radius, while the temperature obtained for the large core mass model M1b declines rapidly towards the center at much larger radius. As was pointed out in §2, for a mass profile with a shape consistent with the lensing observations, the gas temperature should be approximately constant down to small radius in order to produce the required X-ray surface brightness. The solutions M1a and M2 demonstrate that the presence of a cooling flow is indeed consistent with a constant gas temperature over a substantial part of the flow, and with a centrally peaked cluster mass distribution, as required to yield gravitational lensing. Notice, however, that the requirement of having a sufficiently high temperature to explain the typically observed deflection angles for clusters that are approximately spherically symmetric is not satisfied by A478 (one of the most luminous clusters known).

The density distributions $f(r, \rho)$ of the M1a,b solutions at radii $r = (400, 200, 100, 50) \, h^{-1}$ kpc, are presented in Figure 6. The density distribution at $r = 400h^{-1}$ kpc is the one chosen initially to obtain the X-ray profiles in Fig. 4. The distribution chosen for the M2 model is identical to that of the M1a model, and the evolution of the distribution with radius is very similar in the two models. We therefore do not present a similar figure for the M2 model. The density distribution required is wider for the more concentrated mass models.

The approximately constant temperature obtained for the M1a and M2 solutions over the range $r > 20h^{-1}$ kpc is a result of the wide density distribution and of the steep gravitational potential. As the gas flows, its entropy decreases due to the cooling, but if the cooling is not fast enough the temperature may increase as the gas is compressed, owing to the pressure gradient in the cluster. In particular, when the initial density distribution is sufficiently wide, the phases of low density (and high temperature) will cool at a slow rate, and their temperature will increase as they flow inward. The radial dependence of the temperature of various gas phases is presented for the M1a and M1b solutions in Figure 7 (again, the M2 solution is similar to the M1a solution). While all the phases cool as the gas flows inward in the M1b solution, the wider density distribution and larger pressure gradient in the M1a model causes the hot phases to heat up as they flow inward at large radius. As a result, the average temperature is approximately constant in this radial range.

We emphasize here that, in general, a declining temperature profile towards the center will be obtained within the X-ray core of clusters if large mass core models are assumed (see, e.g., the models of Edge & Stewart 1991a). However, *this has nothing to do with the presence of a cooling flow, but it is simply a result of the equation of hydrostatic equilibrium, when the mass profile near the center is assumed to be less concentrated than the gas density profiles determined from X-ray observations.* In order for the gas to be more centrally concentrated than the mass, its velocity dispersion (related to the gas temperature) must drop faster to the center than that of the mass. But, as we have mentioned in § 2, the velocity dispersion of the mass already drops to the center in any model where the slope of the density increases monotonously with radius, up to an isothermal slope. The gas temperature must then decrease even more, and in fact in many cases it must start decreasing *outside* the cooling radius for reasonable cluster ages. However, this depends on the detailed mass model that is used; for the King model, which has a slope $\rho(r) \propto r^{-3}$ at large radius, the temperature only decreases closer to the center.

The $\dot{M}(r)$ profiles of the numerical solutions are plotted in Figure 8a. Their normalization is fixed by choosing a representative value of $2 \cdot 10^{44}$ erg s$^{-1}$ for the luminosity $L_{100}$ of the gas enclosed inside a $100h^{-1}$ kpc sphere. The energy flowing into a sphere of a given radius, due to the mass inflow, is emitted as X-ray luminosity by the gas enclosed in the sphere. If the gravitational energy gained by the gas as it flows inwards is small compared to its thermal energy, then the luminosity $L(r)$ of the gas enclosed in a sphere of radius $r$ is equal to the influx of gas enthalpy at this radius. In this case $L(r)$ is related to the mass flux by

$$\dot{M}_{NG}(r) = \frac{2}{5}\frac{\mu L(r)}{T_M(r)}, \qquad (15)$$

where the subscript $NG$ denotes that no gain of gravitational energy is included. If the gain of gravitational energy is not negligible, then for a given $L(r)$, $\dot{M}(r)$ should be smaller than $\dot{M}_{NG}(r)$, since part of the luminosity is due to the energy gained by the gas as it falls into the gravitational potential. The ratios of $\dot{M}(r)$ to $\dot{M}_{NG}(r)$ for the three solutions are plotted in Figure 8b. At large radius, $r > 250 h^{-1}\,\mathrm{kpc}$, the contribution of gravitational energy to the luminosity is substantial, and is $\sim 25\%$ higher for the concentrated mass models M1a and M2 than for the large core model M1b. For this reason, at large radius the mass inflow rate of the M1a and M2 solutions is $\sim 25\%$ lower than that of the M1b solution. At smaller radius, $r < 150 h^{-1}\,\mathrm{kpc}$, the difference in $\dot{M}$ between the models rises up to a factor of 2 to 3. This difference is not due, however, to gravitational energy gain. The contribution at $r = 100 h^{-1}\,\mathrm{kpc}$ of gravitational energy to the luminosity is small, of order 20%, and also very similar for the various models. Thus, the gas enthalpy flux at this radius is similar for all models. The difference in the mass deposition rates results from the fact that the model with the large mass core has a much lower gas temperature near the center.

To further emphasize and clarify this point, we have numerically solved the multiphase flow equations for the large mass core model M1b, choosing an outer radius $r_0 = 100 h^{-1}\,\mathrm{kpc}$, an outer pressure $p_0 \equiv p(r_0)$ corresponding to a gas temperature of $11\,\mathrm{keV}$ at $r_0$, and an outer density distribution $f_0(\rho)$ that produces an emissivity profile $\epsilon(r)$ similar to those in Figure 4. The numerical mass flow rate, normalized for $L_{100} = 2 \cdot 10^{44}\,\mathrm{erg\,s^{-1}}$, is presented in Figure 8a. The temperature and mass flow rate at $r = 100 h^{-1}\,\mathrm{kpc}$ of this solution are similar to those of the M1a and M2 solutions. This demonstrates that, for a given $L(r)$ and $T(r)$, the mass inflow rate $\dot{M}(r)$ is only weakly dependent on the model assumed for the cluster mass. Note that, for the large mass core model M1b with the high temperature we have assumed, the pressure is almost constant and the temperature should continue to rise for $r > 100 h^{-1}\,\mathrm{kpc}$ in order to produce a sufficiently steep gas density profile, and a steep X-ray profile, with $\epsilon(r) \propto r^{-3}$. The temperature would then be much higher than the virial temperature, which is highly unrealistic.

The contribution of gravitational energy to the luminosity in the steep mass models M1a and M2 is not much higher than in the large core model M1b, since the $\dot{M}$ profiles are also steeper for M1a and M2. Thus, most of the gas moves only over a small distance before it is deposited in the models M1a and M2, and there is not much gain of gravitational energy. The $\dot{M}$ profile is decreasing inwards in the M1a and M2 solutions for $r < 200 h^{-1}\,\mathrm{kpc}$, while it is constant down to $r \sim 100 h^{-1}\,\mathrm{kpc}$ in the M1b solution; for $r < 100 h^{-1}\,\mathrm{kpc}$, the $\dot{M}(r)$ profiles are well approximated by a power law $r^\delta$, with $\delta \approx 1.6$ for the M1a and M2 solutions, and $\delta \approx 1.4$ for the M1b solution.

The steady cooling flow solution is valid at radii smaller than the cooling radius $r_{cool}$, defined as the radius at which the isobaric cooling time,

$$\tau_p = \frac{5}{2}\frac{T}{\mu\rho\Lambda}, \qquad (16)$$

is equal to the cluster age. In addition to the cluster age, the cooling radius also depends on the scaling of the density and temperature of the solution, which may be varied as mentioned at the end of § 3.1. For $h = 0.5$, for example, fixing the luminosity within the central sphere of 200 kpc to $L_{200} = 1.6 \cdot 10^{45}\,\mathrm{erg\,s^{-1}}$, the X-ray temperature to 6.5 keV (which corresponds to the cluster A478), and the age to $2 \cdot 10^{10}\,\mathrm{yr}$, gives cooling radii $r_{cool} \approx 320\,\mathrm{kpc}$ for the M1a and M2 solutions, and $r_{cool} \approx 350\,\mathrm{kpc}$ for the M1b solution. The scaling of $r_{cool}$ with cluster luminosity, age, and X-ray temperature depends on the emissivity and temperature profiles. For the numerical solutions M1a,b and M2, the cooling time is $\tau_p \propto r^{1.8}$ for $r > 80 h^{-1}\,\mathrm{kpc}$. Typically, outside the cooling radius the density profiles are close to isothermal and the temperatures are approximately constant. If we assume $\tau_p \propto r^2$, then using the scalings (12-14), and denoting the cluster age as $\theta_a H_0^{-1}$, we derive the following approximate scaling of the cooling radius,

$$H_0 r_{cool} \propto (H_0^2 L)^{1/4} T^{-3/8} H_0^{-1/4} \theta_a^{1/2}. \qquad (17)$$

Here, $T$ is an average cluster X-ray temperature, and we have assumed $\alpha = 0.5$. Equation (17) shows that the cooling radius does not depend strongly on the assumed Hubble constant value. Furthermore, $r_{cool}$ also depends weakly on the cluster luminosity and temperature scale.

In hierarchical theories, clusters are formed by mergers. The gas in any given cluster should have been cooling since it was last mixed and shock-heated; therefore, the ages of clusters should typically be the time between major mergers. If $\Omega$ is not very small, most present clusters should have undergone a major merger during the past half of the age of the universe. The fraction of clusters that experienced a major merger during the past 20% of the age of the universe increases with $\Omega$ and is $\sim 30\%$ for $\Omega = 1$ (Richstone, Loeb, & Turner 1992; Lacey & Cole 1993). If a typical value for the age of cooling flows is half the age of the universe, this gives $\theta_a \simeq 1/3$ in equation (17), and $r_{cool} \simeq 100 h^{-1}\,\mathrm{kpc}$ for a cluster like A478. In an open universe, cooling radii should generally be slightly larger.

The models M1a and M2 have $\dot{M}$ profiles which continue to grow up to a larger radius than in Model M1b. The difference arises because in the large core solution the temperature is rising outwards, so that although $L(r)$ increases with radius the ratio $L(r)/T(r) \propto \dot{M}(r)$ remains approximately constant for $r > 100 h^{-1}\,\mathrm{kpc}$. For the small core solution, $T(r)$ is approximately constant so that $L(r)/T(r)$ increases rapidly with radius, and a lot more mass is deposited at larger radius. This implies that *the insensitivity of the total mass inflow rate within the cooling radius, $\dot{M}_c \equiv \dot{M}(r = r_{cool})$, to the age (e.g., Edge et al. 1992), is only a property of models with a large core for the mass*, where the $\dot{M}$ profile is already quite flat for the typical values of $r_{cool}$. When the mass is more concentrated, as required by gravitational lensing, then the derived mass deposition rates do depend on the assumed age of the cluster.

Finally, we notice that the emissivity profile $\epsilon(r)$ of the cooling flow solution for the mass model M1a presents a significant flattening at the cooling radius $r_{cool} \approx 100 h^{-1}\,\mathrm{kpc}$, although the mass core radius is only $r_c = 30 h^{-1}\,\mathrm{kpc}$. The form of $\epsilon(r)$ depends upon the gas density distribution. However, the following simple argument shows that, for an isothermal mass model, a substantial flattening of the emissivity profile must always occur at the cooling radius. The emissivity profile outside the cooling radius is proportional to $r^{-4}$ (assuming that the mass profile is isothermal, that the gas is in hydrostatic equilibrium, and that the gas temperature is close to the virial temperature). The luminosity of a shell of radius $r$ and thickness $\Delta r \propto r$ drops, for $\epsilon \propto r^{-4}$, like $r^{-1}$, so that most of the luminosity comes from the innermost part of the gas. If the emissivity profile was not flattened inside the cooling radius, this would imply that most of the gas is cooling and emitting its energy at a radius much smaller than the cooling radius. However, if the gas is not cooling appreciably down to radii small compared to the cooling radius, its temperature would rise substantially due to the adiabatic heating. This rise in temperature would lead to a flattening of the emissivity, since the gas temperature would become higher than the mass virial temperature. In order that most of the energy is not emitted near the center, the density profile must be flatter than $r^{-3/2}$. To demonstrate this we have numerically solved the multiphase flow equations for a mass model of the form (1) with a very small core radius, $r_c = 3h^{-1}$ kpc. We have used for this calculation the same boundary conditions at $r_0 = 400h^{-1}$ kpc that were used for the M1a solution (for which $r_c = 30h^{-1}$ kpc). The results are presented in Figure 9(a,b). As expected, the gas temperature is rising towards the center and $\epsilon(r)$ exhibits a substantial flattening at $r \approx 100h^{-1}$ kpc. Although the core radius is much smaller in this solution than in the M1a solution, the X-ray profile flattens at approximately the same radius.

### 3.3. Analysis of Self-Similar Solutions

Nulsen (1986) discussed multiphase flows with radially self-similar density distributions. For such flows, analytical solutions may be constructed for a family of mass profiles $M(r)$. In this subsection we use these analytical solutions to study the dependence of cooling flow properties on the mass model, in a more general way than we have done above from a few numerical solutions. In order for the results of such an analysis to apply to physically realistic flows, the self-similar solutions should give an approximate description of flows obtained for realistic $M(r)$ and $\epsilon(r)$ profiles. This is indeed the case, as will be shown in § 3.4.

Let us consider a multiphase flow with an invariant density distribution,

$$\psi(r,\rho) = -\dot{M}(r)\, I[\xi(r,\rho)]\,, \tag{18}$$

where

$$\xi(r,\rho) = \rho/\bar{\rho}(r)\,. \tag{19}$$

With this normalization of the shape function $I$ and of the self-similar density variable $\xi$, $I$ must satisfy (from the condition $\int d\rho f(\rho) = 1$ and eq. [5])

$$I(\xi = \infty) = 1, \qquad \int_0^\infty \xi^{-1} \frac{dI}{d\xi} d\xi = 1\,. \tag{20}$$

Using (4) and (6) one finds that (18) is a solution only if $I$ takes the form

$$I(\xi) = \begin{cases} \left[1 - (\xi/\xi_0)^{-(2-\alpha)}\right]^\nu, & \text{if } \xi > \xi_0\ ; \\ 0, & \text{otherwise}\ , \end{cases} \tag{21}$$

or the form

$$I(\xi) = \exp\left[-(\xi/\xi_0)^{-(2-\alpha)}\right]\,. \tag{22}$$

Here, $\nu$ is a free parameter while $\xi_0(\nu,\alpha)$ is determined by (20). Both forms extend to $\rho = \infty$ and have a high density "cooling tail" $f(\rho) \propto \rho^{\alpha-4}$. The density distribution (21) has a cutoff at low densities at $\rho = \xi_0 \bar{\rho}$; the distribution is very narrow for small $\nu$, and wider for large $\nu$ (see Fig. 10). Equation (22) is obtained from (21) in the limit $\nu \to \infty$, so we can restrict our treatment to solutions of the form (21). For these solutions $\xi_0$ is given by

$$\nu^{-1}\xi_0(\nu,\alpha) = B\left(\nu, 1 + \frac{1}{2-\alpha}\right) \equiv \int_0^1 t^{\nu-1}(1-t)^{1/(2-\alpha)}dt\,. \tag{23}$$

The entropy and mass flux equations, (4) and (6), reduce for $I(\xi)$ given by (21) to

$$\frac{d}{dr}\left[\left(\frac{p^{3/5}}{\bar{\rho}}\right)^{\nu(2-\alpha)}\dot{M}^{-1}\right] = 0\,, \tag{24}$$

and

$$\frac{1}{4\pi r^2}\frac{d\dot{M}}{dr} = (2-\alpha)\nu\xi_0^{2-\alpha}\dot{\rho}\,, \tag{25}$$

where we have defined

$$\dot{\rho}(p,\bar{\rho}) \equiv \frac{2}{5}\frac{\Lambda\left(T=\mu p/\bar{\rho}\right)\bar{\rho}^3}{p}\,. \tag{26}$$

The quantity $\dot{\rho}$ is the deposition rate per unit volume that is obtained by taking the emissivity of gas with density $\bar{\rho}$ and pressure $p$, and dividing by the enthalpy per unit mass of the same gas.

Thus, the radial evolution of a multiphase invariant solution is described by the hydrostatic equation (3) and by (24)-(26). For a power law mass profile,

$$M(r) \propto r^{\lambda_M}\,, \tag{27}$$

these equations have analytical solutions where $p$, $\bar{\rho}$ and $\dot{M}$ are also power laws, determined by $\nu$ and $\lambda_M$. It is convenient to express the solution in terms of the mass power $\lambda_M$ and the emissivity power $\lambda_L$, where

$$\epsilon \propto r^{-\lambda_L}\,.$$

The solution then takes the form

$$\bar{\rho} \propto r^{-[\lambda_L + \alpha(\lambda_M - 1)]/2}\,, \tag{28}$$

$$\frac{p}{\bar{\rho}} = \frac{2}{\lambda_L - (2-\alpha)(\lambda_M - 1)}\frac{GM}{r}\,, \tag{29}$$

$$\dot{M} = \frac{5\xi_0^{2-\alpha}}{\lambda_L + (3+\alpha)(\lambda_M - 1)}4\pi r^3\dot{\rho} \propto r^{4-\lambda_M-\lambda_L}\,. \tag{30}$$

The parameter $\nu$ is given by

$$\nu = \frac{5}{2-\alpha}\frac{4-\lambda_L-\lambda_M}{\lambda_L+(3+\alpha)(\lambda_M-1)}\,. \tag{31}$$

The steepest possible density profiles are obtained when $\nu \to 0$. In this case, the density distribution is very narrow and the solution tends to a homogeneous cooling flow (the fraction of the mass deposited in a given range of radii is negligible). For example, for the isothermal mass profile ($\lambda_M = 1$), this leads to $\lambda_L < 3$, which is the condition discussed at the end of § 3.2. Realistically inhomogeneous cooling flows should correspond to large values of $\nu$, and more extended gas profiles; for example, $\lambda_L = 2$ for $\nu = 5/3$ and for the isothermal case, which gives a density profile $\bar{\rho} \propto r^{-1}$ and a constant inflow velocity. This shows that *the effect of a cooling flow when the mass profile is isothermal is to raise the gas temperature to a value higher than the virial temperature, and make the gas more extended than the mass*. The gas is generally more extended for wider density distributions. It is easily seen from the above equations, that the conclusion that the gas is more extended than the mass in a cooling flow remains correct for $\lambda_M < (2+\alpha)/(1+\alpha)$, for these self-similar solutions.

The flattest density profiles are obtained in the limit $\nu \to \infty$, corresponding to the widest self-similar distributions. In this case, according to (24), the "average" entropy ($\bar{S} \propto \log(p\bar{\rho}^{-5/3})$) is constant with radius. Cooling will generally cause the entropy of the gas to decrease as it flows inward; however, the deposition of the cooler phases results in an increase of the average entropy of the remaining phases in the flow, if the cold gas is at some point decoupled from the hot phases and ceases to contribute to the average weight of the gas. The case $\nu \to \infty$ corresponds therefore to the critical case when these two effects are exactly balanced and the entropy is constant. For wider initial density distributions, a larger fraction of the gas is deposited at a given radial range, leading to a decrease of $\bar{S}$ with radius. The solution to the cooling flow is then unstable to global convection. Thus, the density profiles obtained for the widest self-similar distributions ($\nu \to \infty$) are the flattest profiles for which the flow is convectively stable (Nulsen 1986).

The gas temperature in the solutions (28)-(30) follows the virial temperature, $T_{vir}/\mu = GM/r$. Such solutions exist only for

$$\lambda_M < 1 + \frac{\lambda_L}{2-\alpha}, \tag{32}$$

or, equivalently, for

$$\lambda_M < 1 + \frac{3}{(3-\alpha) + (2-\alpha)\nu}. \tag{33}$$

For a given $\lambda_L$ (or a given $\nu$), the ratio between gas and virial temperature, given by (29), increases with $\lambda_M$, i.e., as the mass profile is made shallower. When $\lambda_M$ reaches the maximum value in (32), the gravitational term in the hydrostatic equilibrium equation (3) is no longer important, and the solution becomes isobaric. Basically, the virial temperature of the mass drops to the center too fast to be followed by the temperature of the cooling gas, and the gas temperature stays larger than the virial temperature as the gas flows inward at constant pressure. The invariant solution is then of the form

$$\bar{\rho} \propto r^{-\lambda_L/(2-\alpha)}, \tag{34}$$

$$\dot{M} = \frac{(2-\alpha)\xi_0^{2-\alpha}}{\lambda_L} 4\pi r^3 \dot{\rho} \propto r^{3-\frac{3-\alpha}{2-\alpha}\lambda_L}. \tag{35}$$

For this solution $\nu$ is given by

$$\nu = \frac{3}{\lambda_L} - \frac{3-\alpha}{2-\alpha}. \tag{36}$$

Again, the condition $\nu > 0$ gives a maximum possible steepness for the gas profile, and the flattest stable solutions are obtained for $\nu \to \infty$.

From (31), (32) and (36) we see that a given emissivity profile requires wider distributions (larger $\nu$) for steeper mass profiles. For wider distributions a larger fraction of the gas is deposited at large radii, so that a given emissivity implies steeper $\dot{M}$ profiles for steeper mass profiles (eqs. (30), (35)). These conclusions are in accordance with the numerical results of § 3.2.

Finally, we examine the relation between $\dot{M}(r)$ and $L(r)$, the luminosity within a sphere of radius $r$. For the analytical self-similar solutions the emissivity is given by

$$\varepsilon(r) = \Lambda_0 T_M^\alpha \bar{\rho}^2 \int_{\xi_0}^\infty d\xi\, \xi^{1-\alpha} \frac{dI}{d\xi}, \tag{37}$$

and $\dot{M}$ and $L$ are related by

$$\dot{M}(r) = \eta(\lambda_L, \lambda_M) \frac{2}{5} \frac{\mu L(r)}{T_M(r)}, \tag{38}$$

where

$$\eta(\lambda_L, \lambda_M) = \begin{cases} 1 & \text{for isobaric solutions,} \\ \frac{3-\lambda_L}{3-0.8\lambda_L-0.2(2-\alpha)(\lambda_M-1)} & \text{otherwise,} \end{cases} \tag{39}$$

(the expression in (39) is obtained by using the relation $B(x, y+1)/B(x,y) = y/(x+y)$ for beta functions). The relation (38) is the same as (15), obtained for negligible gravitational work, except for the numerical coefficient $\eta$. The deviation of $\eta$ from 1 is due to the gain of gravitational energy by the infalling gas. This gain is more significant for lower $\eta$ values. $\eta$ decreases with $\lambda_L$ because for a steeper emissivity profile the gas flows through a larger distance toward the center before cooling below X-ray temperatures. This increases the gain of gravitational energy. For a constant $\lambda_L$, $\eta$ increases with $\lambda_M$, implying that the energy gain is larger for steeper mass profiles. However, the increase of $\eta$ with $\lambda_M$ is small and, furthermore, for $\lambda_L$ values typically observed the energy gain is not very large. For example, for $\alpha = 0.5$ and $\lambda_L = 1.5$, $\eta$ changes from 1 to $\sim 0.8$ over the $\lambda_M$ range. This implies that the relation between $\dot{M}$, $L$ and $T$ is almost independent of the mass profile, and that for typical emissivity profiles this relation is approximately given by (15), where gravitational work is neglected.

### 3.4. *Applicability of the Self-Similar Analysis*

Nulsen (1986) showed that the self-similar solution (21) with $\nu = \nu_0$ describes the asymptotic behavior in the limit $r \to 0$ of flows where the initial density distribution has a low density cutoff of the form $f \propto (\rho - \rho_{min})^{\nu_0 - 1}$, for $\rho \to \rho_{min}$. In fact, we shall see that the analytic self-similar solutions have a much wider range of applicability. It is shown below that they also give an approximate description of multiphase cooling flows obtained for realistic mass profiles $M(r)$ and emissivity profiles $\epsilon(r)$, over a substantial part of the radial range.

Let us first consider the numerical solutions of § 3.2 for the mass models M1a,b. In Figure 11 the numerical density distributions obtained at several radii are compared with the self-similar distributions (21) with $\nu = 3.1$ for the small core model M1a and with $\nu = 2$ for the large core model M1b. In both cases, the numerical profiles in the radial range $30h^{-1}$ kpc $< r < 90h^{-1}$ kpc agree well with the self-similar profiles. Furthermore, over the radial range where the flow is approximately self-similar, power-law approximations to the $\dot{M}$ and $\epsilon$ profiles are reasonably good. It has been mentioned in §3.2 that $\dot{M} \propto r^\delta$ for $r < 100h^{-1}$ kpc (see Fig. 8a), and the power law behavior of $\epsilon(r)$ is demonstrated in Figure 12.

This self-similar behavior is due to the fact that the evolution of the densities of different phases in a cooling flow is decoupled. That is to say, the evolution of a particular phase depends on the others only through the global pressure and velocity profiles. Thus, if an initial density distribution $\psi_0(\rho)$ is well approximated by a self-similar distribution of the form (21),

$$\psi_0(\rho) \propto \left[1 - \left(\frac{\rho}{\tilde{\rho}}\right)^{-(2-\alpha)}\right]^{\nu_i}, \qquad (40)$$

over some density range $\rho_a < \rho < \rho_b$ (where $\tilde{\rho} < \rho_a$ is only a fitting parameter and is not necessarily related to a true cutoff of the density distribution), and if most of the gas is initially in phases within this density range, then the density distribution will also be well approximated by the self-similar solution at all other radii, except near the center, where the phases with initial densities smaller than $\rho_a$ are left. Phases with densities outside the range $\rho_a < \rho < \rho_b$ affect this evolution only through the pressure profile $p(r)$.

The examples presented above demonstrate that the analytic self-similar multiphase flow solutions can be used to approximately describe flows obtained for realistic mass profiles and emissivity profiles, which are not power laws but can be approximated as such over some radial range. The approximate self-similar description holds for most of the radial range over which the gas is deposited. Thus, the results obtained in the previous subsection for the analytical self-similar solutions are expected to be generally valid for physically relevant cooling flows.

## 4. CONSEQUENCES FOR CLUSTER EVOLUTION

We have found that the constraints on the form of the mass profile obtained from gravitational lensing, which basically require the mass to be more concentrated than it was usually assumed previously, are perfectly consistent with observations of the X-ray profiles in clusters of galaxies, and the presence of cooling flows, although in general the X-ray temperatures are small compared to the ones that should correspond to the observed deflection angles. In fact, the presence of a cooling flow in a potential well dominated by dark matter, where the mass profile is not very flat, will generally cause the gas to be *hotter* than the dark matter near the center, and consequently to be more extended than the dark matter. We shall now examine the consequences implied by an increased temperature and a more extended density profile within the cooling radii of galaxy clusters for the evolution of galaxy clusters with redshift.

Kaiser (1986, 1991) showed how the evolution of the luminosity and temperature functions of X-ray clusters can be predicted in self-similar models. The basic idea is that, in the absence of any physical process giving a characteristic scale for clusters of galaxies, the density and temperature profiles of clusters in a hierarchical model ought to be self-similar, so that clusters should have the same properties at all epochs except for a rescaling of the typical scale of collapse. We shall summarize the relations derived in such self-similar models, and then see how they can be altered by the presence of cooling flows.

The self-similarity is an exact property in cosmological models with no global curvature or vacuum density and a power spectrum equal to a pure power-law, and when radiative cooling of the gas, as well as heating by mechanisms other than the shocks caused by gravitational collapse, are ignored. In this case, there is absolutely no preferred scale for non-relativistic clusters. In general, predictions for the power spectrum based on assuming scale-invariant primordial fluctuations (such as the CDM model) contain a characteristic scale, related to the observable horizon at the epoch when matter and radiation have equal densities. However, the change of slope of the power spectrum is quite gentle in CDM, and it can still be approximated as a power-law over a limited range of scales. The rms amplitude of density fluctuations smoothed over a comoving scale $R$ can then be written as

$$\delta\rho/\rho \propto R^{-(3+n)/2} \ . \tag{41}$$

For CDM, $n \simeq -1$ on the cluster scales at the present time. Typically, clusters at a scale $R$ will form when the linearly extrapolated density fluctuations are of order unity. In the linear regime, density fluctuations grow linearly with the scale factor $a$, and therefore the characteristic scale of the clusters at a given epoch is $R \propto a^{2/(3+n)}$. The velocity dispersion of the clusters is $\sigma \propto (aR)/t$, where the age of the universe is $t \propto a^{3/2}$. Thus, the cluster temperature is $T \propto \sigma^2 \propto a^{(1-n)/(3+n)}$. At the same time, the cluster comoving number density will vary as $N(a) \propto R^{-3} \propto a^{-6/(3+n)}$.

In the absence of cooling, the X-ray luminosity will be $L \propto \rho M T^\alpha$, where $\rho \propto a^{-3}$ is the characteristic density, $M \propto R^3$ is the mass, and we have parameterized the cooling function as $\Lambda \propto T^\alpha$ on the relevant range of temperatures. Then, following Kaiser's notation, we have the following relations for the variations of temperature, luminosity and number density for any class of self-similar clusters at different epochs:

$$\Delta \log T = \frac{1-n}{3+n} \Delta \log a \ ; \tag{42a}$$

$$\Delta \log L = -\frac{3 + 3n - \alpha(1-n)}{3+n} \Delta \log a \ ; \tag{42b}$$

$$\Delta \log N = -\frac{6}{3+n} \Delta \log a \ . \tag{42c}$$

These relations produce the evolution of the luminosity function of clusters that is shown in Figure 3 of Kaiser (1991), for $\alpha = 1/2$. The problem with these relations is that, for $n = -1$ or higher, the number of high luminosity clusters is predicted to be higher in the past, whereas the opposite is observed (Edge et al. 1990; Gioia et al. 1990; Henry et al. 1991). For $n = -2$, the evolution of the luminosities is in better agreement with observations, but the temperatures are predicted to be much lower in the past. The origin of this problem is that the core radii of clusters are predicted to decrease with redshift, proportionally to the scale radius $R$. The higher density of the gas then causes a higher luminosity in the past.

Now, let us consider how these relations should be changed due to cooling. As we have seen, for concentrated mass profiles consistent with lensing the effect of a cooling flow will be to flatten the gas density profile within the cooling radius, so that the core radius of the gas is of the order of the cooling radius. The cooling radii of clusters will increase towards the past, owing to the higher densities, and this will reduce their luminosities. We assume that the gas profile is isothermal outside the cooling radius, $r_c$, and is effectively flat within $r_c$ (i.e., most of the luminosity is emitted near the radius $r_c$). Then, the cluster luminosity has an additional dependence on the ratio of the cooling radius to the proper scale of the cluster:

$$L \propto a^{-3} R^3 T^\alpha \, (Ra/r_c) \ . \tag{43}$$

The cooling radius is determined by the condition that the cooling time at $r_c$ should be equal to the cluster age. This is proportional to the age of the universe (it must be born in mind here that more massive clusters, with higher $T$, have generally merged more recently at a fixed epoch [e.g., Lacey & Cole 1993], but this does not alter the self-similarity with $a$): $T^{1-\alpha}/\rho(r_c) \propto a^{3/2}$, where $\rho(r_c) \propto a^{-3}(Ra/r_c)^2$ is the density at $r_c$, and we assume again an isothermal profile outside $r_c$. This gives

$$\frac{r_c}{aR} \propto a^{-3/4} T^{-(1-\alpha)/2} . \qquad (44)$$

Substituting in equation (43), we obtain:

$$L \propto a^{-9/4} R^3 T^{(1+\alpha)/2} . \qquad (45)$$

Using $R \propto (Ta)^{1/2}$, we find that the luminosity should depend on the cluster temperature and epoch as $L \propto a^{-3/4} T^{2+\alpha/2}$. If we include the dependence of the characteristic cluster temperature on epoch from (42a), we find

$$\Delta \log L = -\frac{1 + 11n - 2\alpha(1-n)}{4(3+n)} \log a . \qquad (46)$$

The evolution predicted by this new equation is shown in Figure 13, where we have rescaled the present luminosity function according to Edge et al. (1990) using equation (46), and (42c) for the number density of clusters, to a redshift $z = 0.2$, for values of $n$ going from $-2$ to $1$. This can be compared with Figure 3 in Kaiser (1991), which is equivalent to our Figure 13, but using the relations (42b,c). When the effects of cooling are included, the abundance of X-ray clusters can decline at the highest observed luminosities for $n = -1$ or less. It is not clear if this is enough to explain the observed negative evolution of the high luminosity X-ray clusters, but it certainly goes a long way towards reconciling them with the observations. The exact evolutionary rate is very sensitive to the precise form of the cluster luminosity function at the high luminosity end. The lower luminosity clusters, however, should still be more abundant in the past.

The hypothesis of self-similarity used to derive the rate of cluster evolution states that the population of clusters at different epochs differ only by a rescaling of the typical cluster mass. We can also make the stronger assumption that, at a fixed epoch, the density profiles of clusters are all self-similar, and they differ only by one parameter which gives their mass, or velocity dispersion. This has been proposed by Navarro et al. (1994), who find from numerical simulations that the mass density profiles of clusters can indeed be approximately described by a universal function, when appropriately rescaled. Since clusters are continuously merging, there must be at least another variable property of clusters of galaxies, which is the degree of substructure. This must depend on the merging history of the cluster. However, all clusters that are approximately relaxed could have similar density profiles, and they should obey a luminosity-temperature relation at a fixed epoch which, in the absence of cooling, is $L \propto R^3 T^\alpha \propto T^{3/2+\alpha}$. When cooling is included, then from equation (45) this is modified to $L \propto T^{2+\alpha/2}$. This is still substantially shallower than the observed relationship, $L \propto T^3$ (e.g., Edge & Stewart 1991a). This may be caused by a difference in the cluster density profiles depending on temperature, or also a difference in the average degree of substructure in clusters depending on temperature. In fact, since higher temperature clusters have typically merged more recently, they should have in average more substructure and, in the ones where the center is relaxed, smaller cooling radii owing to the younger average ages. However, there may also be another effect due to cooling which we have not included here: the cooling causes the temperature in the cluster cores to increase, in order to support the gas with a shallower density profile. This will result in an increase of the global, radially averaged cluster temperature, which will be more important in low temperature clusters given their larger cooling radius. Thus, the range of cluster temperatures is reduced, and the luminosity-temperature relation should be steeper.

The predicted evolution of the core radius of clusters, assumed to be equal to the cooling radius, is found from (44) to be $r_c \propto a^{3/4} T^{\alpha/2}$. The core radii are predicted to be smaller in the past, although they evolve less fast than in the case when cooling is not included.

## 5. ON THE MORPHOLOGICAL CLASSIFICATION OF CLUSTERS

The classifications of clusters of galaxies that have been proposed (see Bahcall 1977 and van den Bergh 1977 for reviews) are generally two-dimensional. One parameter is associated with cluster richness, velocity dispersion, X-ray luminosity and temperature. These properties all correlate with each other, although with a substantial amount of scatter. The second parameter is related to structural properties (Bautz & Morgan 1970; Rood & Sastry 1971) and the fraction of ellipticals (Oemler 1974). Type I clusters are regular and centrally condensed around a large central galaxy, with most of the galaxies being ellipticals and S0's. Type III clusters do not have a dominant galaxy, are irregular and contain more spirals. This second parameter has usually been ascribed to a difference in dynamical evolution, with denser clusters having evolved further in a fixed elapsed time, and corresponding to Type I clusters, while clusters evolving more slowly would correspond to the other types (e.g., Hausman & Ostriker 1978). Jones & Forman (1984) proposed another cluster classification. Based on X-ray observations, they separated clusters into an X-ray dominant class (XD), having small X-ray core radius with the emission centered on a giant galaxy, and non X-ray dominant (nXD), which have larger core radius and no dominant central galaxy.

As we have mentioned above, the hypothesis that the cluster density profiles produced by hierarchical gravitational collapse are approximately self-similar (see Navarro et al. 1994) leads to a classification of clusters containing only two parameters. One is the scale of the cluster, which should correlate with cluster mass (and total optical luminosity if the mass-to-light ratios are not highly variable in different clusters), the galaxy velocity dispersion and the X-ray temperature.

The second parameter is the degree of substructure. Even if all clusters acquire similar density profiles when they relax after each successive merger, they should still differ depending on their recent merging history. Here, we propose that both second parameters of the optical and X-ray classifications (Type I/III and XD/nXD) correspond to the degree of substructure. Clusters which have not had a major merger during several dynamical times, and have undergone only slow accretion of matter in the form of small clumps compared to the total cluster mass, should be relaxed and could all have similar density profiles. The distribution of mass in these clusters ought to be regular and in dynamical equilibrium, and the gas should also be in hydrostatic equilibrium around the center of the dark matter halo. A giant galaxy should typically be at the center, formed from mergers of the galaxies that resided in the dark matter halos which merged to form the cluster, and also subsequent dynamical friction of other galaxies (Ostriker & Tremaine 1975; Merritt 1985). These can correspond to the Type I and XD clusters, which have peaked X-ray profiles, and therefore are found to have cooling flows. On the other hand, clusters which are in the process of merging from various subclusters of similar mass should have multiple mass clumps originating from the centers of the merging units. These clusters should have no dominant galaxy, but instead several large ellipticals, which will still be surrounded by dark matter halos being disrupted as they orbit through the cluster. The galaxy distribution can appear more irregular. Spiral galaxies may be more abundant, because the matter that was previously in filamentary structures connecting the merging halos should contain gas-rich galaxies, and they should be merging rapidly into the newly formed cluster. Since there is no stationary deep potential well, the X-ray gas can have a very large core. In some of these clusters, a strong shock in the gas may produce an irregular X-ray image, but the gas should not necessarily have an irregular distribution in all non-relaxed clusters. Owing to its collisional nature, the gas cannot follow the motions of the dark matter clumps, and it has to wait for the dark matter to settle on a stationary equilibrium before it can collapse on a concentrated, deep potential well.

We therefore propose that clusters in the nXD class defined by Jones & Forman have all had a recent merger, and that their large X-ray cores do not reflect a large core in the potential well, but simply the presence of substructure. This makes several clear predictions: first, clusters with large X-ray cores should not have a central dominant galaxy, but they should have several giant ellipticals accompanied by swarms of galaxies around them. The large X-ray cores should also not be completely flat, but they should show signs of substructure when observed at a sufficiently high signal-to-noise. The archetypical example of an nXD cluster with no cooling flow, the Coma cluster, does indeed have two giant ellipticals near the center of similar luminosity (which we interpret as the remaining cores of two subclusters that have been merging), and shows signs of substructure in the central X-ray isophotes (White, Briel, & Henry 1993).

Owing to their large X-ray cores, clusters in the nXD class are classified as not having cooling flows. Edge et al. (1992) have also suggested that these clusters may all be the result of recent mergers, and they interpret their large core radii as a result of the disruption of the cooling flow during the merger. In this paper, we have shown instead that if we consider a young cluster where cooling is unimportant, but where the gas is already in hydrostatic equilibrium in a stationary potential well that is able to cause the observed gravitational lenses, and the gas follows the mass initially, we should expect the gas to be *more* concentrated than in an older cluster where the central gas has cooled. Such a young cluster would therefore also be classified as having a cooling flow, given the usual condition that the central cooling time is less than the Hubble time. According to our interpretation of the XD/nXD classification of clusters, the *cause* of the large gas cores in nXD clusters is central substructure in the mass distribution, and the consequent lack of hydrostatic equilibrium for the gas, rather than the absence of a cooling flow; and the *cause* of the small X-ray cores in XD clusters is the concentrated mass distribution in the mass profiles of relaxed clusters, rather than the presence of a cooling flow.

If the large X-ray cores of the nXD clusters were instead due to the presence of large cores in the mass distribution of relaxed clusters, this would imply that density profiles should vary from cluster to cluster, with some clusters being more extended than others. This is not supported by the simulations of Navarro et al. (1994), and if true it would provide an important constraint for theories of cluster formation. The preponderance of central substructure in clusters with large X-ray cores can also be tested with gravitational lensing observations: a uniform, large core would yield a central region with no observable shear, whereas a cluster with central substructure should produce a complicated shear pattern indicating many mass clumps. Substructure is indeed obviously present in many lensing clusters (see, e.g., Pelló et al. 1991, 1992; Miralda-Escudé 1993; Kneib et al. 1993), but it is not clear if it is more abundant in clusters with large X-ray cores.

We have seen that X-ray profiles should generally flatten at the radius where the cooling time is of order of the age of the cluster, owing to the effects of cooling. However, if a merging object can deposit a large amount of dense, cool gas near the center of the cluster, this could lead to a short lived cooling flow where the cooling time is much shorter, and therefore the core radius could be much smaller. At later times, the gas flowing to the center should be hotter and the X-ray core should increase. This could have taken place in the Perseus cluster, which has a very steep central X-ray profile (Branduardi-Raymont et al. 1981). The unusually bright optical line-emission nebula in the center of this cluster (Heckman et al. 1989) may also be related to a temporarily large deposition rate.

## 6. CONCLUSIONS

Observations of gravitational lensing provide evidence that the mass density profiles of clusters do not have large cores compared to the radius where arcs are observed. In addition, they give a value for the total mass of the cluster within the radius of the arc, which can be compared to the mass needed to support the X-ray gas in hydrostatic equilibrium.

We have found that the constraints on the *form* of the mass profile, which basically require the mass to be more concentrated than it was usually assumed previously, are perfectly consistent with observations of the X-ray profiles in clusters of galaxies and the presence of cooling flows. In fact, the presence of a cooling flow in a potential well dominated by dark matter, where the mass profile is not very flat, will generally cause the gas to be *hotter* than the dark matter near the center, and consequently to be more extended than the dark matter. On the other hand, the typical cluster X-ray temperatures seem to be generally too low to account for the observed deflection angles in lensing clusters, when hydrostatic equilibrium and spherical symmetry is assumed; this problem was found by Miralda-Escudé & Babul (1995) in particular cases where both lensing observations and X-ray temperatures were available. It is not clear at present if this originates from a peculiar orientation or structure of the clusters producing gravitational lensing, or from a more general physical process which modifies the dynamical equilibrium of the gas.

We find that the gas temperatures in the cooling regions of clusters of galaxies should be nearly constant, contrary to what has generally been expected so far. This is not necessarily contradicted by the observation of a cooler X-ray temperature near the cluster center (e.g., Schwarz et al. 1992; Allen et al. 1993; Allen & Fabian 1994), since this may simply reflect an increase of the width of the temperature distribution, and of the amount of cool gas due to the presence of a cooling flow. The average temperature of all phases, which determines the pressure, is generally higher than the emission-weighted temperature because the emission is proportional to the square of the density, and it depends on the instrumental sensitivity as a function of frequency. Recent observations with the ASCA satellite of the Centaurus cluster (Fukazawa et al. 1994) show evidence for a multiphase medium, and they do not indicate a large temperature decrease near the center.

A clear prediction of a deep gravitational potential well in clusters is the presence of hot gas near the center. This was shown in Figure 7. For the model M1a, the temperature of the hottest phases increases toward the center in the cooling region, and this balances the increasing fraction of the gas at low temperatures. This ought to result in an excess of high energy photons in the X-ray spectrum in the cluster center. However, the increase of the highest temperatures is not very substantial, and the excess of high energy photons is below current upper limits (see Allen et al. 1992).

The contribution of gravitational energy gain to the gas energy is potentially more important for the more concentrated mass models. However, it was shown that the total mass deposition rate is determined primarily by the emissivity profile and the temperature at the cooling radius, and depends only weakly on the assumed mass profile if these two quantities are fixed. This is a result of the fact that the gravitational work done on the inflowing gas is small and does not vary greatly between models, so that the flow rate is determined by the gas temperature and the energy emitted in the cooling region. Concentrated mass models also have a stronger dependence of the total mass deposition rate on the assumed cluster age, compared to models with large mass cores. The age of a cluster should probably be identified with the time since the last major merger, when the gas in the center was shocked for the last time; this time should vary from cluster to cluster, and its median value should be shorter in a flat universe than in an open universe. In a large mass core model, an older age simply implies that the gas is cooling from a larger radius and from a higher temperature, with a narrow initial density distribution. In contrast, when the mass core is small, the density distribution is wider and the gas temperature is constant, so as the cluster ages more mass is deposited at large radius.

This suggests that the mass deposition rates inferred from the X-ray spectrum of clusters (Fabian et al. 1991 and references therein) might be used as a probe of the distribution of ages of clusters of galaxies, by comparing them to the values obtained from X-ray imaging. However, there are several uncertainties in making such a comparison. First, if the initial temperature distribution in the intracluster medium is wide, then all the emission from the cluster will have a complicated mixture of emission from different temperatures. This could be solved once spatially resolved spectra are obtained. But even in this case, the abundance of cool gas could be affected by thermal conduction fronts around deposited clouds. Cold clouds could fall through the potential well of the cluster and radiate much more energy in the process. This could be emitted in soft X-rays if most of the gravitational energy was released on a bow shock near the cloud surface, causing an increase in the spectral values of $\dot{M}$.

Cooling flow models with deep potential wells can explain the observations in a more natural way than models with large mass cores. The observed X-ray cores are readily explained from the effects of cooling, and they do not need to be attributed to a special scale in the mass profiles. Since cooling occurred in the past over a larger fraction of the gas in clusters, the core radii should be larger in the past *relative* to the virial radii of halos, and this changes the X-ray luminosity evolution rate expected in self-similar hierarchical models. We have shown that the inclusion of the effects of cooling implies that the predicted evolution of X-ray clusters at the highest observed luminosities is negative for $n \lesssim -0.5$ (Fig. 13). For $n = -1$, the abundance of high luminosity clusters declines from the present to $z = 0.2$ by a factor of $\sim 1.5$. The lower luminosity clusters, however, should still be more abundant in the past. Some clusters have very large X-ray cores that cannot be explained if the clusters are in equilibrium and the mass profiles do not have large cores. We have proposed that these clusters are not in dynamical equilibrium, and this offers an interpretation of the morphological classification of clusters which we have described in § 5.

Throughout this paper, we have only analyzed the standard model of the central cooling regions of clusters of galaxies, where the cooling gas is deposited and eventually forms low-mass stars, while the overlying gas flows inwards (e.g., Fabian 1994). In order for such a cooling flow to form, one needs to assume that there is no energy source which can balance the energy lost by the gas through radiative cooling. If such an energy source was present, cold clouds formed from the cooling gas could simply reevaporate into the hot gas. Thus, the occurrence of cooling flows in clusters of galaxies has not yet been demonstrated, since we have no proof that such an energy source is absent and we have no evidence for an average inward flow of the multiphase gas. The recent discovery of X-ray absorbing material in clusters of galaxies (White et al. 1991) may help to clarify the fate of the cooling gas. We have seen that the most simple cooling flow models, where all the gas phases are comoving, are in good agreement with the observations of the intracluster medium when the mass profiles are sufficiently concentrated to produce gravitational lensing. Alternative models where an energy source is present are difficult to test, since there is no clear prescription for the rate at which heat could be injected in different gas phases at different radii, and the relative motions of the phases. Other ways to test such alternative models will be investigated in future research.

**Acknowledgment:** EW and JM thank the W. M. Keck Foundation for financial support. EW is also supported by NSF grant PHY 92-45317.

– 28 –

# FIGURE CAPTIONS

**Fig. 1:** Mass density profiles are shown for Model 1 [eq. (1)] in Fig. 1a and Model 2 [eq. (2)] in Fig. 1b as solid lines, with the core radii indicated, which are consistent with lensing. Dashed lines are gas density profiles for isothermal gas and the three indicated values of $\beta \equiv \mu\sigma^2/T$.

**Fig. 2:** X-ray surface brightness for the gas profiles in Fig. 1 for $\beta = 1$, compared with the ROSAT PSPC data for A478 (Allen et al. 1993). The solid line is the convolution of the X-ray profile of Model 2 with a gaussian with FWHM $= 33h^{-1}$ kpc ($30''$). The convolution has a negligible effect (not shown for Model 1).

**Fig. 3:** Mass models used for the numerical cooling flow solutions. Models M1a and M1b are of the form (1) with $\sigma = 1325 \, \mathrm{km \, s^{-1}}$, and $r_c = (30, 125)h^{-1}$ kpc, respectively. The M2 model is of the form (2) with $r_c = 265h^{-1}$ kpc and $\sigma = 1305 \, \mathrm{km \, s^{-1}}$. All models include a central galaxy, as described in the text.

**Fig. 4:** Surface brightness profiles of the numerical cooling flow solutions for the various mass models, compared with the ROSAT PSPC data for A478. Thick lines show the X-ray profiles convolved with a gaussian with FWHM $= 33h^{-1}$ kpc.

**Fig. 5:** Temperature profiles of cooling flow solutions. Thick lines show the mass averaged temperature, and narrow lines the emissivity averaged temperature. Line types indicate the same models as in Fig. 4.

**Fig. 6:** Density distribution functions of models M1a and M1b at four different radii. The initial distribution was chosen at $r = 400h^{-1}$ kpc. Notice that mass is deposited in the M1a solution for $r < 200h^{-1}$ kpc, and only for $r < 100h^{-1}$ kpc in the M1b solution.

**Fig. 7:** Evolution of the temperature of eight gas phases with radius, for the M1a and M1b solutions. The initial temperatures $T_i$ of the phases are such that the fraction of the initial gas inflow from temperatures $T > T_i$ constitute $(100, 80, 60, 40, 20, 10, 5, 1)\%$ of the total inflow.

**Fig. 8a:** Mass inflow rates of the M1a, M1b and M2 solutions, normalized to a luminosity $2 \cdot 10^{44} \, \mathrm{erg \, s^{-1}}$ within a sphere of $100h^{-1}$ kpc. The long-dash line is for the M1b mass model, but with a high initial temperature, $T = 10$ keV, at $r = 100h^{-1}$ kpc, which gives an X-ray profile similar to those of the other solutions for $r < 100h^{-1}$ kpc (normalized to the same luminosity within $100h^{-1}$ kpc).

**Fig. 8b:** Ratio of mass inflow rate $\dot{M}$ to the rate $\dot{M}_{NG}$ required to produce the same luminosity with the same gas temperature, if the gravitational energy gain of the cooling gas is negligible. This ratio is equivalent to the ratio of gas enthalpy inflow at a given radius to the luminosity emitted by the gas at smaller radii. Line types for different models are as in Fig. 8a.

**Fig. 9a:** Surface brightness profile of a solution for Model 1 for the mass profile with $r_c = 3h^{-1}$ kpc (long-dash line), compared with the same solution with $r_c = 30h^{-1}$ kpc (Model M1a; solid line).

**Fig. 9b:** Mass-weighted temperature for the same two models as in Fig. 9a.

**Fig. 10:** The self-similar cumulative density distribution for various $\nu$ values.

**Fig. 11:** Cumulative density distributions of the M1a and M1b models at various radii, compared with self-similar density distributions.

**Fig. 12:** Power law fits to the emissivity profiles of the numerical models M1a and M1b, for $r < 100h^{-1}$ kpc. The break at larger radius is due to a discontinuity in the derivative of the initial density distributions (see Fig. 6).

**Fig. 13:** Evolution of the cluster X-ray luminosity function from the present one (taken from Edge et al. 1990) to a redshift $z = 0.2$, as predicted by a self-similar hierarchical model incorporating the effects of cooling flows, for various values of the power spectrum slope $n$. $n = -2$ corresponds to the leftmost curve at the high-luminosity end, and $n = 1$ to the rightmost curve.